\documentclass[prl,twocolumn,superscriptaddress,nopacs,amsmath,amssymb,nofootinbib]{revtex4-2}
\usepackage[utf8]{inputenc}
\usepackage{mathptmx}
\usepackage[T1]{fontenc}
\usepackage{float}
\usepackage{amssymb}
\usepackage{graphicx} 
\usepackage{esint}

\usepackage{hyperref} 
\usepackage[nameinlink]{cleveref}
\crefname{subsection}{subsection}{subsections}

\usepackage{color}
\usepackage{booktabs}
\usepackage{makecell}
\usepackage{soul}
\usepackage{sidecap} 
\usepackage{gensymb} 

\usepackage{enumitem}
\setlist[itemize]{leftmargin=*}

\renewcommand\[{\begin{equation}}
\renewcommand\]{\end{equation}} 

\makeatother

\usepackage{babel}
\usepackage{physics}

\makeatletter
\newcommand*\bigcdot{\mathpalette\bigcdot@{.5}}
\newcommand*\bigcdot@[2]{\mathbin{\vcenter{\hbox{\scalebox{#2}{$\m@th#1\bullet$}}}}}
\makeatother

\usepackage{mathtools}
\usepackage{scalerel}

\makeatletter
\DeclareFontFamily{OMX}{MnSymbolE}{}
\DeclareSymbolFont{MnLargeSymbols}{OMX}{MnSymbolE}{m}{n}
\SetSymbolFont{MnLargeSymbols}{bold}{OMX}{MnSymbolE}{b}{n}
\DeclareFontShape{OMX}{MnSymbolE}{m}{n}{
    <-6>  MnSymbolE5
   <6-7>  MnSymbolE6
   <7-8>  MnSymbolE7
   <8-9>  MnSymbolE8
   <9-10> MnSymbolE9
  <10-12> MnSymbolE10
  <12->   MnSymbolE12
}{}
\DeclareFontShape{OMX}{MnSymbolE}{b}{n}{
    <-6>  MnSymbolE-Bold5
   <6-7>  MnSymbolE-Bold6
   <7-8>  MnSymbolE-Bold7
   <8-9>  MnSymbolE-Bold8
   <9-10> MnSymbolE-Bold9
  <10-12> MnSymbolE-Bold10
  <12->   MnSymbolE-Bold12
}{}

\let\llangle\@undefined
\let\rrangle\@undefined
\DeclareMathDelimiter{\llangle}{\mathopen}%
                     {MnLargeSymbols}{'164}{MnLargeSymbols}{'164}
\DeclareMathDelimiter{\rrangle}{\mathclose}%
                     {MnLargeSymbols}{'171}{MnLargeSymbols}{'171}
\makeatother

\renewcommand{\footnoterule}{\kern -1ex\rule{\linewidth}{0.5pt}\\\vspace{1ex}}

\usepackage[makeroom]{cancel} 

\usepackage{titlesec}
\titlespacing{\subsection}{0pt}{\baselineskip}{0.5\baselineskip}

\begin{document}

\title{Mechanism of charge transfer and electrostatic field fluctuations in high entropy metallic alloys}
\author{Wai-Ga D. Ho}
\affiliation{Department of Physics and National High Magnetic Field Laboratory, Florida State University, Tallahassee, FL, USA}
\author{Wasim Raja Mondal}
\affiliation{Department of Physics and Astronomy, Middle Tennessee State University, Murfreesboro, Tennessee 37132, USA}
\author{Hanna Terletska}
\affiliation{Department of Physics and Astronomy, Middle Tennessee State University, Murfreesboro, Tennessee 37132, USA}
\author{Ka-Ming Tam}
\affiliation{Department of Physics and Astronomy, Louisiana State University, Baton Rouge, LA 70803, USA}
\author{Mariia Karabin}
\affiliation{National Center for Computational Sciences, Oak Ridge National Laboratory, Oak Ridge, Tennessee 37831, USA}
\author{Markus Eisenbach}
\affiliation{National Center for Computational Sciences, Oak Ridge National Laboratory, Oak Ridge, Tennessee 37831, USA}
\author{Yang Wang}
\affiliation{Pittsburgh Supercomputing Center, Carnegie Mellon University, Pittsburgh, Pennsylvania 15213, USA}
\author{Vladimir Dobrosavljevi\'{c}}
\affiliation{Department of Physics and National High Magnetic Field Laboratory, Florida State University, Tallahassee, FL, USA}
\begin{abstract}
High entropy alloys present a new class of disordered metals which hold promising prospects for the next generation of materials and technology. However, much of the basic physics underlying these robust, multifunctional materials – and those of other, more generic forms of disordered matter – still remain the subject of ongoing inquiry. We thus present a minimal-working model that describes the disorder-driven fluctuations in the electronic charge distributions and electrostatic "Madelung" fields in disordered metals. Our theory follows a standard perturbative scheme and captures the leading contributions from dominant electronic processes, including electrostatic screening and impurity scattering events. We show here that a modest first-order treatment incorporating these effects is sufficient to reproduce the linear charge transfer trends featured in both high-entropy and other conventional alloys, our model also shedding light on the microscopic origins of these statistical features. We further elaborate on the nature of these electronic charge and Madelung field fluctuations by determining how these emerge from the statistics of the underlying disorder, and how these can be described using the linear response formulation that we develop here.  In doing so, our work answers various questions which have long-perplexed the disordered materials community. It also opens up possible avenues for providing systematic corrections to modern first-principles approaches to disorder-modeling (e.g. the conventional CPA method) which currently lack these statistical features.

\end{abstract}
\maketitle

\section{INTRODUCTION} 


In condensed matter sciences, the presence of disorder in real materials can often have profound and enriching effects on their mechanical, chemical, and electromagnetic properties, as well as on the physics underlying these details.  With most ordinary solids comprising regular crystalline arrangements of atoms, further incorporating disorder adds a new element of randomness into their makeup. And whether this occurs through arbitrary reassignment of different atoms to random positions on a crystal lattice (pure chemical disorder), randomization/distortion of the lattice positions themselves (structural disorder), or some combination of the two, the effects can be quite dramatic, exerting appreciable influence over material properties in the solid state. 

For instance, in both the arts of metallic alloying and ceramic firing -- millenial practices both dating back well into prehistoric times -- disordered amalgamations consisting of multiple chemical elements are formed, often possessing both enhanced and useful qualities. These include, for example, \linebreak materials presenting with augmented strength and hardness, or more generally, those which benefit from increased resistance to various mechanical and/or thermochemical forms of stress and degradation. Moreover, with various ingredients and refined processing techniques involved synergistically in their production, many of these disordered solids exhibit properties which have shown to be quite versatile and tuneable (e.g. by changing chemical ratios).  Exploiting these details, humans throughout recorded history have been able create much of the materials they needed to serve as crucial components in both primitive and modern forms of technology -- the development of which has tracked closely behind innovations made in materials processing and engineering.


This tradition continues today, with major fronts of contemporary research dedicated towards the study of a new class of disordered metals, known as high-entropy alloys%
\footnote{As an aside, we note that parallel research efforts have also developed in the study of high-entropy \textit{ceramics} -- these being the inorganic, nonmetallic counterparts to the alloys we describe here. The main focus of this work, however, shall be placed on metals whose electrons are sufficiently mobile; this will allow us to draw more intuition from traditional problems dealing with similarly dilute Fermi systems and borrow from the standard ideas and techniques which are used in their treatment. Henceforth therefore, we shall restrict our discussion to high-entropy alloys, perhaps reserving the topic of ceramic compounds for future work.}, %
as well as towards realizing applications for these in the next generation of materials and technology. Unlike traditional alloys -- which, for thousands of years, have been manufactured by taking a host of one or two principal elements and imbuing this with other substances in minor concentrations -- high-entropy alloys generally comprise $\gtrsim 4$ principal elements which are mixed in considerable (often close to equiatomic) concentrations. And though the field is still quite young -- the world's first *official* introduction to high-entropy alloys occurring not 20 years prior \cite{HEA_Yeh2004,cantor2004} -- these materials have sparked large and growing research interest, thanks in part to some of the remarkable properties they have showcased in the years following their inception. 

Beyond extraordinary displays of mechanical strength and resistance to corrosive/oxidizing agents, exceeding those levels of even the most popular conventional alloys used today,  other prominent examples have also demonstrated highly efficient/stable electromagnetic and optical responses, biocompatibility, and even superconductivity \cite{HEA_Yeh2004,cantor2004,HEA_SC,li_mechanical_2019,george_high-entropy_2019,zhou2023}. Hence, much of the technological prospects awaiting these high-entropy alloys lie in their potential to serve as resilient, multifunctional materials with hopeful applications in future communications, biomedical, energy storage, and aerospace industries (to name a few) \cite{MuSTpaper2022,george_high-entropy_2019}. Naturally so, 
\linebreak 
the race to develop these compounds, optimize their design, and capitalize on their useful qualities (in a commercial viable manner) has taken precedence in recent years. 

With the past two decades seeing fruitful research efforts and exciting discoveries regarding these materials' empirical properties, much has also been learned about the underlying connection between their unique structural qualities and their multi-principal compositions. On a basic level, during the alloy's fabrication process, standard thermodynamics tells us that both the entropy $\Delta S_{\textrm{mix}}$ and enthalpy $\Delta H_{\textrm{mix}}$ of mixing (that is, of the mixed alloy with respect to those of its pure ingredients) contend with one another to determine the spontaneity of the mixing process and the stability of the mixed product itself. For a given temperature $T$, this competing process is captured through the Gibbs free energy of mixing $\Delta G_{\textrm{mix}} = \Delta H_{\textrm{mix}} - T \Delta S_{\textrm{mix}}$. And it is in these high-entropy alloys that both the diversity and the comparable abundance of their constituent elements result in a large availability of possible structural configurations, the subsequent enhancement of entropy (hence the nomenclature), and its out-performance of the enthalpy term. In the end, this all works towards the stabilization ($\Delta G_{\textrm{mix}}<0$) of the alloy's solid-state structure, which is often seen adopting a single crystalline phase when pure \cite{HEA_Yeh2004,cantor2004,phaseselection2019,george_high-entropy_2019}. 


Despite the latest advances in the field however, there remain a number of open issues regarding more specific details on the internal structure and properties of these high-entropy alloys, as well as those of disordered solids in general -- this limiting, to some degree, our ability to more effectively design and produce what's needed for modern technological applications. A more comprehensive understanding of the relevant physics in these compounds is therefore required to bring them into the next stage of their development. Fortunately for us, recent times have also seen major advancements made in the quantum theory of solids, both in terms of our understanding and our ability to employ it for the purposes of solving wider and increasingly complex problems, these naturally lending themselves towards the furthering of the abovementioned goals. \linebreak \vspace{-\baselineskip}

Progress along these lines have seen innovative solutions for traditional obstacles faced by the disordered materials community -- in large part, difficulties in taking standard methods proven successful in treating ordinary, periodic solids (i.e. those based on density-functional theory \cite{HKthrm,kohnSham1965}), and adapting these for purposes of treating the disorder problem.  See, unlike in periodically ordered solids, where precise crystal translational invariance greatly reduces the necessary physical considerations down to just a single (or maybe a handful) of smaller unit cells, the characteristic randomness of disordered solids is thorough%
\footnote{As opposed to shallow -- i.e. in cases of weak doping, or where we otherwise have a host substance that bears a minor/diffuse concentration of impurities. This situation is often reasonably accommodated by using density-functional, or other similar band theory methods, to treat an analogous problem that \textit{is} periodic. (references? silicon/semiconductor doping problems?)}, %
spanning large length scales over which structural and/or chemical defects are pervasive and command a leading role in the pertinent physics. And to properly account for these, the general implication is that the standard density-functional approach would require massive supercell treatments and have to meet impractically high computational demands in order to accomplish even the most routine calculations. As such, further steps were necessarily taken before accurate disorder modeling could become a more feasible reality.

The last four decades or so have seen most of the enduring work done on this front, the fruits of these labors providing us today with a number of available options for simulating and studying the properties of disordered solids, all to varying degrees of approximation, accuracy, and computational affordability \cite{kkrCpaChargeSC,sqs,lsms1,lsms2}. This has culminated in recent work \cite{MuSTpaper2022}, which submits some of the most popular of these methods to benchmark testing and performance review in the especially relevant context of high-entropy alloys. Those that \cite{MuSTpaper2022} addresses can broadly be understood to fall within two main classes: 
\begin{itemize}[nosep,leftmargin=*]
    \item \textit{effective medium theories} -- instead of treating the disordered crystal directly, these methods traditionally involve solving an analogous problem in which we remove each of the solid's elemental/constituent species, embedding these instead in an effective medium that is determined by the atomic species' average environment. This local mean-field mapping to a single-site problem is afforded through what's commonly known as the coherent-potential approximation (CPA) \cite{CPA,KKRCPA1,KKRCPA2,kkrCpaChargeSC}. And while this does grant conceptual simplicity and computational affordability, such convenience comes at the high cost of sacrificing key physical details; as implied by our description, this includes the discardment of disorder-driven fluctuations in each atom's local chemical environment, and hence the (intraspecies) identities of the atoms themselves.
    \item   \textit{supercell methods} -- these are generally more faithful in their representation of the long-ranged disorder. For example, in the locally self-consistent multiple scattering (LSMS) method \cite{lsms1,lsms2}, a large $\sim \! \mathcal{O}[1000]$-atom supercell is treated by establishing a radial cutoff around each atom, the effects beyond which are then ignored. With the aid of parallel computing architecture, this radial truncation cuts the computational cost down to an expensive (compared to CPA) yet manageable price. Moreover, with the local environments surrounding around each atom now retained in the density-functional step, this preserves to a reasonable degree the intraspecies statistics arising from disorder, thus yielding a more physically satisfying description of the solid.%
\end{itemize}

Now what is also important to note is that, from the LSMS calculations presented in this work, it was found that the disorder-driven statistics for the:
\begin{itemize}[nosep,leftmargin=*]
    \item \textit{electronic charge-transfer} -- that is, the net gain/loss of electrons experienced by each atom upon its incorporation into the solid
    \item \textit{electrostatic "Madelung" fields} -- i.e. the local electrostatic potential that is felt by a given atom and is generated by the remaining bulk of the solid
\end{itemize}
\noindent these both demonstrated much of the same qualitative behaviors and trends as those established in earlier works for more conventional alloys \cite{oldAlloys1,oldAlloys2,oldAlloys3}. Included in these were Gaussian distributions for the intraspecies statistics of either independently considered quantity, as well as a statistically linear "qV" relationship shared jointly between them. To note, these results appear to suggest the existence of universal mechanisms which underlie the charge transfer and Madelung field statistics within the broader class of disordered metals. However, the details surrounding these remain yet unclear, and there persist various open questions regarding what specifically governs these features (e.g. qV trendline parameters, their dependence on various physical quantities, etc.) and how we can most simply understand and describe their associated statistics. 

The goal of this work is to provide answers to these questions, which we shall soon develop by first constructing a minimal-working model that captures the basic qualities common to all chemically disordered alloys%
\footnote{As is the case in \cite{MuSTpaper2022} and other relevant works \cite{oldAlloys1,oldAlloys2,oldAlloys3}, we shall restrict our considerations to the problem of pure chemical disorder. While we note that structural defects as well as other deviations from ideal crystallinity (e.g. secondary, tertiary, amorphous phases) do certainly play a key role in alloy technology -- these being especially relevant when discussing the microstructure which directly impacts aspects of mechanical strength, hardness, etc. -- we maintain that the pure, single-crystal phase presents a more fundamental problem, and that a stronger understanding of this situation holds higher precedence.}.
Using our simplified theory, we will then show that this can reproduce all the same statistical features demonstrated previously by LSMS, and investigate things further to advance our understanding of their nature and origins. From this, we may finally establish what governs these (LSMS) statistics and trends to arrive at the answers we seek.   

As an introductory endnote, we further motivate our work by pointing out that, in \cite{MuSTpaper2022}, it was also shown how the LSMS-provided qV trends can be used to modify CPA's effective medium, providing systematic corrections which enhance the accuracy of CPA's output results (e.g. energetics, effective charge transfers and Madelung potentials, etc.) while preserving its speed and efficiency. Thus, our work -- which seeks to clarify and describe these (LSMS) statistics -- presents an opportunity to develop a methodology which improves upon the traditional CPA approach by furnishing these qV trends in a manner which avoids the computational demands of LSMS.

%


\section{THEORETICAL FRAMEWORK} \label{sec:theory}

\subsection{A disordered Hartree model for random alloys} \label{subsec:DHmodel}

For a simplified description of the chemically disordered alloy, we consider just a single-band model of spinless fermions which occupy an $N$-site lattice with moderate impurity disorder. Our full Hamiltonian is given by
\begin{equation}
    \hat{H} = \hat{H}_0 + \hat{V}
    \qquad ; \qquad 
    \left\{
    \begin{aligned}
        \\[-0.35cm]
        \hat{H}_0 
        & = \sum_{\vb{k} } \xi_{\vb{k}} \,  \hat{c}^{\dagger}_{\vb{k}}\hat{c}_{\vb{k}}
        \\
        \hat{V} 
        & = \sum_i (\epsilon_i + \phi_i ) \, \hat{c}^{\dagger}_i \hat{c}_i 
    \end{aligned}
    \right.
    \label{eq:H}
\end{equation}
and consists of two main parts:

\begin{enumerate}[nosep, leftmargin=*]
    \item a bare term $\hat{H}_0$, which is trivially diagonal in momentum $\vb{k}$-space; it captures nearest-neighbor tight-binding kinetics%
    \footnote{Although our choice to use this tight-binding representation is made rather arbitrarily here, both for its working simplicity and its demonstrative usefulness, we note that much of the work described in this paper is not particularly sensitive to finer band structure details.} %
    with tunneling amplitude $t$. Taking $\vb{a} \in \{\vb{a}_1,\vb{a}_2,...,\vb{a}_d\}$ as the primitive lattice vectors of a generic $d$-dimensional crystal, our one-band dispersion relation is then given by $\xi_{\vb{k}} = -2 \, t \sum_{\vb{a}} \cos[\vb{k}\boldsymbol{\bigcdot}\vb{a}]$.
    \item local perturbations in $\hat{V}$, containing:
    \begin{itemize}[nosep, leftmargin=*]
        \item random site-energies $\epsilon_i$ assigned to each $(i \in N)^{\textrm{th}}$ position in the crystal; these correspond to the disordered background of impurities.
        \item on-site potentials $\phi_i$ which are generated by inter-electron Coulomb repulsions; these are treated at the Hartree mean-field level, where only the average distribution of electrons are accounted for in each site's electrostatic environment. Within such a picture, this Hartree mean-field is constructed for a given site $i$ by summing over the Coulomb contributions from the (net average) charges of the remaining $(j \neq i)^{\textrm{th}}$ sites of the crystal. This prescription coincides with that of the Madelung potential introduced previously, and with $\vb{r}_{i}$ denoting the position of site $i$, we have 
        \begin{align}
            \phi_i 
                & = \sum_{j\neq i} V^C_{ij} (n_j - \langle n \rangle)
                \qquad ; \qquad 
                V^C_{ij} = \frac{1}{|\vb{r}_i - \vb{r}_j|},
                \label{eq:MadPot}
        \end{align}
        where $V^C_{ij}$ represents the intersite Coulomb potential while $\langle n \rangle$ here denotes a positive uniform background which we include to enforce charge neutrality across the entire solid.  
    \end{itemize}
\end{enumerate}

\subsection{General perturbative approach} \label{subsec:PTgen} 

Using standard perturbation theory, we may then relate the electronic profiles of the bare ($\hat{H}_0$) and the full ($\hat{H}$) problems by constructing their (appropriately subscripted and retarded) Green's functions
\begin{equation}
    \hat{G}_{(0)} = (\omega^+ - \hat{H}_{(0)} )^{-1}
    \qquad ; \qquad
    \omega^+ = \lim_{\eta \rightarrow 0^+} (\omega + i\eta)
    \label{eq:GF}
\end{equation}

\noindent and using the Dyson equation
\begin{equation}
    \hat{G} = \hat{G}_0 + \hat{G}_0 \hat{V }  \hat{G}_0 + \hat{G}_0 \hat{V}  \hat{G}_0 \hat{V}  \hat{G}_0 + \cdots
\end{equation}
to express $\hat{G}$ as a series in $\hat{G}_0$ and $\hat{V}$.

Now the local densities of states can be obtained from the local Green's functions (i.e. the on-site matrix elements of $\hat{G}_{(0)}$ above) by extracting their imaginary parts. And integrating these up to the chemical potential $\mu[\langle n\rangle]$ will further provide us with the local charge densities. For either the bare or the full problem, we have
\begin{equation}
    n_{(0)i} = \langle \hat{c}^{\dagger}_i \hat{c}_i \rangle_{(0)}   
    = -\frac{1}{\pi} \int_\infty^\mu \textrm{Im}[\hat{G}_{(0)}]_{ii} \, \dd \omega,
\end{equation}
where the angled braces around the number operator $\hat{n}_i = \hat{c}^{\dagger}_i \hat{c}_i$ are understood to represent its quantum average. Thus, operating with $-(1/\pi)\int_\infty^\mu \textrm{Im}[ \quad]_{ii} \, \dd \omega$ on equation (\ref{eq:GF}) above, we obtain a formula expressing the charge density of the full problem $n_i$ in terms of the charge density of the bare one $n_{0i} = \langle n \rangle$, the latter coinciding with the filling factor thanks to the uniform nature of  $\hat{H}_0$'s solution. From this relationship, we can then express how the local charge density of the bare problem is modified by perturbations to produce the full solution. Such local charge corrections, these being analogous to the charge transfer quantity of interest, are given by \linebreak \vspace{-\baselineskip}
\begin{equation}
    {\delta n}_{i} = n_i - \langle n \rangle = 
    A \widetilde{\epsilon}_i + \sum_{j \neq i} B_{ij} \widetilde{\epsilon}_j + \mathcal{O}[\widetilde{\epsilon}^2],
    \label{eq:DH_eqn1}
\end{equation}
where we note that the (first-order) expansion coefficients
\begin{align}
    A & = -\frac{1}{\pi} \int_{-\infty}^{\mu}\textrm{Im} [{G_0}^2_{ii} ] \, \dd \omega ,
    \label{eq:coefficientAquantum}
    \\
    B_{ij} & = - \frac{1}{\pi} \int_{-\infty}^{\mu}\textrm{Im}[ {G_0}_{ij}^2] \, \dd \omega 
    \quad (\textrm{for $i \neq j$}) ,
    \label{eq:coefficientBijquantum}
\end{align}
often collectively referred to as the Lindhard function, depend exclusively upon the features of the bare model. These have been separated above into their local ($A$) and nonlocal ($B_{ij}$) contributions. Also, in (\ref{eq:DH_eqn1}) above, we have further introduced 
\begin{equation}
    \widetilde{\epsilon}_i = \epsilon_i + \phi_i
    \label{eq:epsTilde}
\end{equation}
as the total on-site perturbation, representing $\hat{V}$'s eigenvalues.

\subsection{Linearizing the problem and achieving self-consistency} \label{subsec:linSC}

To recover the linear qV trends characterizing the joint statistics between charge transfer $\delta n$ and Madelung field $\phi$, it is sufficient to truncate our expansion (\ref{eq:DH_eqn1}) to first-order in $\widetilde{\epsilon}$s. \linebreak Together with our original definition for the Madelung potential (\ref{eq:MadPot}) -- which we now notice contains a $\delta n_j$ nested within its sum -- 
this provides us with a pair of equations which must be solved in a self-consistent manner. 

In our linearized theory, self-consistency is straight\-forward to achieve by Fourier transforming our system of equations -- to which we add (\ref{eq:epsTilde}) -- over to momentum space. Here, the convolution theorem may reduce some of these down to a more algebraic form
\begin{equation}
    \left.
    \begin{aligned}
        {\delta n}_{i}   & = A \widetilde{\epsilon}_i + \sum_{j\neq i} B_{ij} \widetilde{\epsilon}_j 
        \\[0.1cm]
        \phi_i  & = \sum_{j \neq i} V^C_{ij} \delta n_j  
        \\[0.1cm]
        \widetilde{\epsilon}_i & = \epsilon_i + \phi_i 
    \end{aligned}
    \right\}
    \quad
    \longrightarrow 
    \quad
    \left\{
    \begin{aligned}
        \delta n_{\vb{k}} & =  A \widetilde{\epsilon}_{\vb{k}} + B_{\vb{k}} \widetilde{\epsilon}_{\vb{k}}   \vphantom{ \sum_{j\neq i}  }
        \\[0.1cm]
        \phi_{\vb{k}} & = V^C_{\vb{k}} \delta n_{\vb{k}}
        \vphantom{ \sum_{j \neq i} V^C_{ij} \delta n_j  }
        \\[0.1cm]
        \widetilde{\epsilon}_{\vb{k}} & = \epsilon_{\vb{k}} + \phi
        _{\vb{k}}    
    \end{aligned}
    \right.
    .
    \hspace{-0.1cm}
    \label{eq:FT_sysEqns}
\end{equation}
Their $\vb{k}$-space solution is then given by
\begin{align}
    \delta n_{\vb{k}}
     & = M_{\vb{k}} \epsilon_{\vb{k}}
    & \quad   ;  \quad &&   
    M_{\vb{k}} 
    & = \frac{A+B_{\vb{k}} }{1- (A+B_{\vb{k}}) V_{\vb{k}}^C },
    \label{eq:δnk}
    \\
    \phi_{\vb{k}}
    & = \widetilde{M}_{\vb{k}} \epsilon_{\vb{k}}
    & \quad  ;  \quad && 
    \widetilde{M}_{\vb{k}} 
    & = V^C_{\vb{k}} M_{\vb{k}} =  \frac{A + B_{\vb{k}}}{\frac{1}{V^C_{\vb{k}}}- (A+B_{\vb{k}}) } , 
    \label{eq:φk}    
\end{align}
or, returning to site-space (inverting the convolution theorem),
\begin{align}
    \delta n_i & = \sum_j M_{ij} \epsilon_j,
    \label{eq:δni}
    \\
    \phi_i & = \sum_j \widetilde{M}_{ij} \epsilon_j.
    \label{eq:φi}
\end{align}
By inspection, the prior result reveals that $M_{ij}=\partial n_i / \partial \epsilon_j $ and $\widetilde{M}_{ij}= \partial \phi_i / \partial \epsilon_j $; these $M$ and $\widetilde{M}$ objects introduced here shall therefore be understood to describe the (linear) response of the charge density $n=\langle n \rangle + \delta n$ and Madelung fields $\phi$, respectively, to the disordered background of impurities  $\epsilon$.

\subsection{Model features discernible at the single-impurity level}

\begin{figure}[hbt!]
    \centering
    \includegraphics[trim={0.3cm 0.5cm -0.2cm 0.7cm},origin=c,width=0.975\columnwidth]{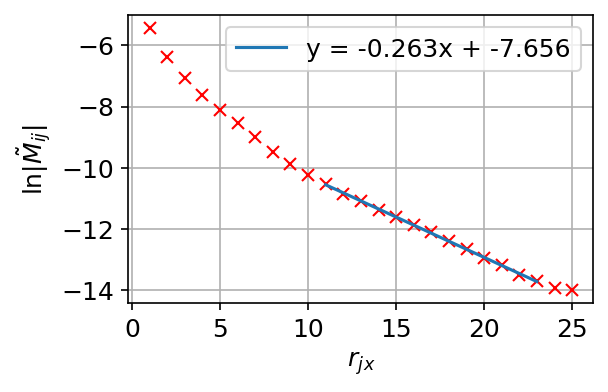}
    \caption{Natural logarithm of the classical Madelung response $\ln| M_{ij} |$, plotted as a function of distance $r_{jx}$ away from an impurity in the half-filled $|\vb{a}|=1$ cubic model ($i \equiv \textrm{origin} $). It is sufficiently linear far from the impurity ($r_{jx} \gtrsim 10$), though some minor flattening does become apparent towards the edge of the cube; this ensures that the real-space profile respects the appropriate boundary conditions (number of sites $N=51\times 51 \times 51$ in this calculation). Temperature $T = 45$ is also set here, which renders the screening length to be $\ell = 3.78... \simeq 1/0.263$ lattice spacings, in good agreement with the least squares fit shown for $\ln|M_{ij}|$'s linear portion.}
    \vspace{-0.4cm}
    \label{fig:classicalMij} 
\end{figure}

All further details concerning our theoretical approach and implementation are reserved for the Supplementary Materials. Proceeding now though, we discuss the physical content of our theory by mentioning that our linearized treatment of the disordered Hartree model retains the following features:  
\begin{enumerate}[nosep,leftmargin=*]
    \item \textit{Friedel oscillations generated by impurity defects} --
    these can be isolated in the non-interacting limit of our theory, where $\{V^C, \phi, \widetilde{M} \} \rightarrow 0$ and $\widetilde{\epsilon} \rightarrow \epsilon $. Local charge corrections for site $i$ due to a single impurity placed on site $j$ are given then in this limit by (compare with equations (\ref{eq:DH_eqn1}),\ref{eq:δni}))
    \begin{align}
        \delta n_i = M_{ij} \epsilon_j  \qquad  ;  \qquad M_{ij} = A \delta_{ij} + B_{ij}.
        \label{eq:deltani_singleimp}
    \end{align}
    These are shown to oscillate and decay away from the impurity in \Cref{fig:friedel} below which uses the half-filled cubic model as a prototypical example. 
    Note that such behavior is analogous to the so-called Friedel oscillations produced around a single impurity scatterer embedded in an ideal Fermi gas; in this related problem, 
    the standard result is that $\delta n_i$ will decay away from $\epsilon_j$ as an inverse cube while oscillating with twice the Fermi wavevector $k_F = 2\sqrt{\mu}$, i.e.
    \begin{align}
         \delta n_i , M_{ij} 
         \sim   
         \frac{ \cos{ \left[ 2 k_F |\vb{r}_{i} - \vb{r}_{j}| \right]  } }{ |\vb{r}_{i} - \vb{r}_{j}|^3 } + \mathcal{O}(|\vb{r}_i - \vb{r}_j|^{-4} ).
    \end{align}
    Following appropriate modifications and some additional effort, one can recover the above result within our first-order disordered Hartree framework, though we refer to \cref{app:FriedelOsc} for more on these details.
    \begin{figure}[hbt!]
        \centering
        \includegraphics[trim={0.3cm 1cm -0.2cm 0.6cm},origin=c,width=1\columnwidth]{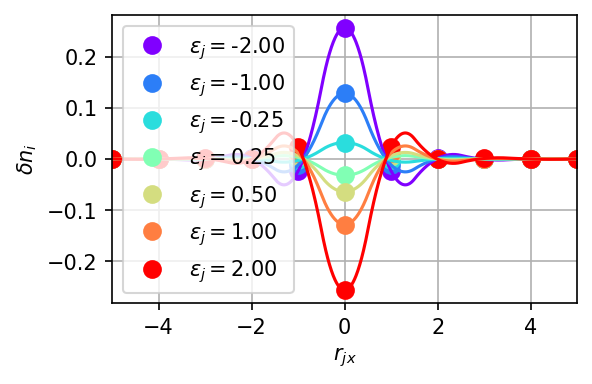} 
        \hspace{0.2cm}
        \caption{ Electronic charge density corrections $\delta n_i$ for site $i \equiv \textrm{origin}$ in the half-filled cubic lattice with unit hopping and lattice spacing ($t=|\vb{a}|=1$). These are generated by an impurity located on site $j$, \linebreak and here we scan the impurity site's position $\vb{r}_{\!j}$ along the $x$-axis. Variations in $\delta n_i(r_{\!jx})$ are then plotted for several choices of impurity strength $\epsilon_j$ across the handful of curves shown. Corrections on-site ($r_{jx} = 0$) are clearly most pronounced, but these become quickly suppressed by about $r_{jx} \simeq 5$ lattice spacings away from $\epsilon_j$.}     
        \label{fig:friedel}
    \end{figure}

    \item \textit{Electrostatic screening of distant impurities} -- this becomes especially transparent in the high-temperature limit of our theory, where we discard quantum hopping as well as any quantities associated with such itinerancy, $\{ t, B \} \rightarrow 0$. \linebreak Thermal fluctuations dominate this regime, these being governed by classical statistics where $A \rightarrow \beta \langle n \rangle (\langle n \rangle - 1) $ and $\beta = 1/T$ is the inverse temperature\footnote{See also \cref{app:classicalDH}. }. The electrostatic Madelung field produced on site $i$ by an impurity $\epsilon_j$ is then given classically as
    \begin{align}
        \phi_i = \widetilde{M}_{ij} \epsilon_j
        \qquad ; \qquad
        \widetilde{M}_{ij}
        = 
        \frac{1}{N} \! \sum_{ \vb{k} }  
        \underbrace{ 
        \frac{A}{\frac{1}{V^C_{\vb{k}}} + A} 
        }_{ 
        \mathclap{ \hspace{1.2cm} \widetilde{M}_{\vb{k}} \textrm{ (classical)} }
        } \, e^{i \vb{k} \bigcdot { (\vb{r}_i - \vb{r}_j) } }.
        \label{eq:φi_classicalSingleImp}
    \end{align}
    And with straightforward numerics, like those provided in \Cref{fig:classicalMij}, we find that the natural logarithm of $\widetilde{M}_{ij}$ becomes sufficiently linear for large distances ${r}_{jx}$ away from the impurity -- a result which indicates an exponential decay in the Madelung response. That is, perhaps denoting some weaker function of impurity distance as $f$, we find that
    \begin{equation}
        \widetilde{M}_{ij} \simeq f(|\vb{r}_{i} - \vb{r}_j|) \, e^{-|\vb{r}_{i} - \vb{r}_j|/\ell}
    \end{equation}
    becomes drastically suppressed beyond a threshold set by the screening length $\ell (= \! \sqrt{-|\vb{a}|^3/(4\pi A)}$ in the classical cubic problem). Hence, we determine that the Madelung field $\phi_i$  responds to an impurity $\epsilon_j$ by electrostatically screening it in a manner reminiscent of the Yukawa form ($f \rightarrow 1/|\vb{r}_i - \vb{r}_j| $). This is most easily understood by considering an idealized continuum problem where, in three dimensions, $V^C_{\vb{k}} \sim 1/|\vb{k}|^2$; in this case also, it becomes immediately apparent that $\widetilde{M}_{ij}$ in (\ref{eq:φi_classicalSingleImp}) reduces to the standard Fourier integral representation of the Yukawa potential. 
\end{enumerate}

\subsection{Disordered Hartree statistics for multiple impurities}

Now we note that, while we did identify and discuss both this Coulomb screening and these Friedel oscillations primarily in the context of single-impurity examples, the extension to multiple impurities is a rather straightforward one to make. In short, our linearized framework is an additive one. And with the density corrections and screened Coulomb potentials from a single impurity given by $M_{ij} \epsilon_j$ and $\widetilde{M}_{ij} \epsilon_j$, \linebreak respectively, one needs simply to sum over many such contributions when considering multiple impurity situations. This will produce the total charge transfer $\delta n_i$ and Madelung field $\phi_i$, as prescribed by formulas (\ref{eq:δni}) and (\ref{eq:φi}) given prior.

\begin{figure}[hbt!]
    \centering
    \hspace{-0.12cm}
    \includegraphics[trim={0.3cm 0.2cm 0 0.8cm}, clip, width=1.005\linewidth]{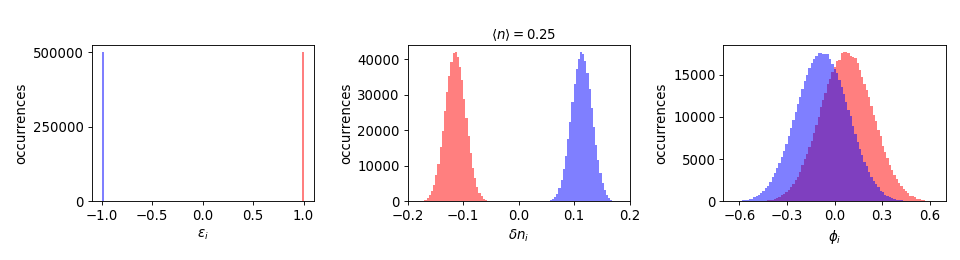}
    \\
    \includegraphics[trim={0.3cm 0.2cm 0 0.8cm}, clip, width=1.015\linewidth]{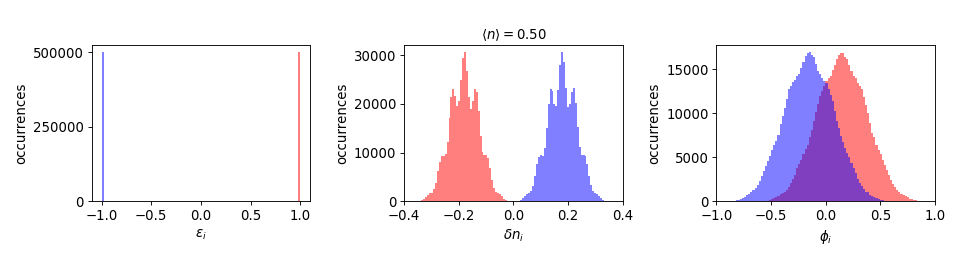}
    \\
    \begin{tabular}{c c}
        \includegraphics[trim={0.3cm 0.6cm 0 0.8cm}, clip, width=0.35\linewidth]{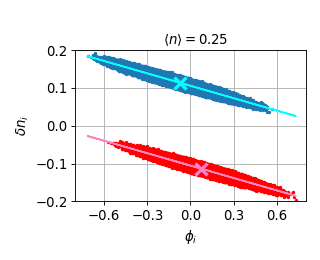}
        &
        \includegraphics[trim={0.3cm 0.6cm 0 0.8cm}, clip, width=0.35\linewidth]{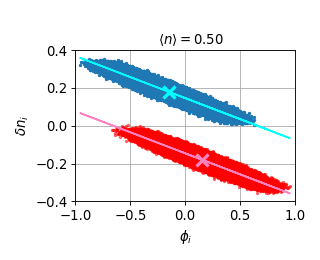}
    \end{tabular}    
    \caption{Local disordered Hartree statistics are sampled for site $i\equiv$ origin in the (top row) quarter- and (middle row) half-filled cubic models, both with equiprobable binary disorder of strength $w = |\epsilon_{j}| = 1$. These 100-bin histograms for the (left column) local charge correction $\delta n_i$, (middle column) impurity potential $\epsilon_i$, and (right column) Madelung field $\phi_i$ all share a converged sample size of $N_r = 10^6$ disorder instances.  In the bottom row, the joint $(\phi_i,\delta n_i)$ data points are then combined and scatterplotted for both fillings, revealing their linear qV relationships, the associated means (crosses) and least squares fit lines also being displayed. All plots in this figure generally distinguish between the positive and negative impurity species by using red and blue shades of coloring, respectively.}
    \label{fig:DHstats}
\end{figure}

With our theory providing the recipe for constructing the $M$ and $\widetilde{M}$ response functions, the multi-impurity problem and procurement of disorder statistics may thus be dealt with in the following manner: for a given input of random impurities $(\epsilon_1,\epsilon_2,...,\epsilon_j,...,\epsilon_N)$, we apply the aforementioned equations and compute $\phi_i$ and $\delta n_i$. Repeating this process, say $N_r$ times, we then collect the local statistics for each instance of disorder -- e.g. in \Cref{fig:DHstats} below for both the quarter- and half-filled cubic models with equiprobable binary disorder. 

At quarter-filling, we note that the screening is sufficiently weak, and that both $\delta n_i = \sum_j M_{ij} \epsilon_j$ and $\phi_i = \sum_j \widetilde{M}_{ij} \epsilon_j$ may comprise a considerable sum of (random) terms. The central limit theorem may then impose Gaussian statistics on either quantity, as demonstrated in the two rightmost panels of \Cref{fig:DHstats}'s top row. Comparing with the half-filled example in \Cref{fig:DHstats}'s middle row, we find that, all else being equal, an increased carrier concentration produces a stronger screening effect which better-shields local quantities from influences beyond their closest-neighboring sites. The statistics of these may then develop more model-/parameter-dependent features, though some broader Gaussian envelope can often remain apparent as is the case here at half-filling. 

\clearpage
\onecolumngrid

\twocolumngrid

As we show next in \Cref{fig:DHstats}'s bottom row, the individual statistics of both $\phi_i$ and $\delta n_i$ may be merged into a combined data set -- the resulting joint statistics, once segregated by impurity species, then scatterplotted to reveal a linear statistical relationship shared between these two quantities. Thus, we verify that our first-order treatment of the simplified disordered Hartree model can adequately recover the linear qV trends relating the random fluctuations of the local charge transfer to those of the Madelung fields. This, together with the Gaussian features made especially apparent in the weakly screened quarter-filled example, is all in good qualitative agreement with the LSMS results obtained in prior studies on both conventional and high-entropy alloys \cite{MuSTpaper2022,oldAlloys1,oldAlloys2,oldAlloys3}. Yet we reiterate here that what remains to be determined is: what governs these (LSMS) statistics and how can we most simply capture them in a quantitatively accurate manner? 

Uncertainties surrounding this broader question are the source of decades-old questions which remain yet unanswered despite innumerable efforts having been spent in this area. Accordingly, in the forthcoming \cref{sec:DHstats}, we shall determine how such statistics can be understood within our simplified framework and try to address some of the most basic questions we can, including: what controls the statistical features underlying these qV trends? Their positions in $(\phi_i,\delta n_i)$ parameter space? Their spreads? Their slopes? Their distribution around the qV trendline? What sort of dependencies do all of these qualities have? To answer these, we will now develop a language which describes the statistics of the impurity $\epsilon$s themselves and will study how these propagate through our disordered Hartree calculations to understand and characterize the various histograms and scatterplots that result.  

\section{ANALYSIS OF STATISTICS IN THE LINEAR DISORDERED HARTREE FRAMEWORK} \label{sec:DHstats}

\subsection{Overall statistics} \label{subsec:DHstats_overall}

For an overall description of our theory's disorder-driven statistics, we first consider impurity disorder that is generally both conserving and truly random. With the ensemble average over all $N_r$ disorder instances denoted by double brackets $\llangle \rrangle$, these two conditions are captured by the following pair of equations
\begin{align}
    \left \llangle \epsilon_j \right \rrangle
    & = 0,
    \label{eq:<εj>}
    \\
     \left \llangle  \epsilon_j \epsilon_{j'}  \right \rrangle
     & = w^2 \delta_{jj'},
    \label{eq:<εjεj'>}
\end{align}
where we introduce $w^2=\llangle \epsilon_j^2 \rrangle$ to gauge the disorder strength. Note that the above encompasses the equiprobable binary case presented in \Cref{fig:DHstats}, where we additionally stipulate that $\epsilon_j = \pm w$ with 50/50 odds. Now the consequences of these for the statistics of our local charge corrections (\ref{eq:δni}) and Madelung fields (\ref{eq:φi}) become immediately apparent, $\delta n_i$ and $\phi_i$'s ensemble averages respectively given by%
\footnote{Note that $M$ and $\widetilde{M}$ are constructed using only the bare profile and lattice structure as prescribed in (\ref{eq:δnk}) and (\ref{eq:φk}). They are therefore completely oblivious to any particular configurational details associated with the disorder, and may be brought out of the expectation value $\llangle \rrangle$.
}
\begin{align}
    \left \llangle  \delta n_i  \right \rrangle
    & =  
    \sum_{j} M_{ij} \left  \llangle \epsilon_j \right  \rrangle   = 0,
    \label{eq:δni_avgall}
    \\
    \left  \llangle \phi_i \right \rrangle
    & 
    =  \sum_{j} \widetilde{M}_{ij} \left  \llangle \epsilon_j \right \rrangle = 0.
    \label{eq:φi_avgall}
\end{align}
Hence, by design, the fluctuations in the charge and Madelung field distributions are centered at zero and are thus globally conserving. Next, measuring the statistical spread about these zero-means, we compute the mean squares as follows 
\begin{align}
    \left  \llangle    \delta n_i^2  \right \rrangle     
    & = \sum_{jj' }  M_{ij} M_{ij'} \left  \llangle \epsilon_j \epsilon_{j'} \right \rrangle
    = w^2 \sum_{j} M_{ij}^2,
   \label{eq:δni^2_avgall}
    \\
    \left  \llangle   \phi_i^2  \right \rrangle
    & =  \sum_{jj' }  \widetilde{M}_{ij}  \widetilde{M}_{ij'}   \left  \llangle     \epsilon_j \epsilon_{j'} \right \rrangle
    = w^2 \sum_{j} \widetilde{M}_{ij}^2.
    \label{eq:φi^2_avgall}
\end{align}

\subsection{Species-resolved statistics} \label{subsec:DHstats_byspecies}

Now to distinguish finer chemical features in our disordered Hartree statistics, we define a species-resolved average $\llangle \rrangle_{i \in \alpha}$ to be taken over the ($N_\alpha < N_r$)-sized sample subset that has been filtered for instances where $\epsilon_i$ belongs to species $\alpha$. This bias fixes the local average $\llangle \epsilon_j \rrangle_{i \in \alpha}$ (provided $i=j$), while the intersite correlator remains almost oblivious to the filter.
\begin{align}
    \left \llangle \epsilon_j \right \rrangle_{i \in \alpha}  
    & = \epsilon_{\alpha} \delta_{ij} , 
    \label{eq:<εj>α}
    \\
    \left \llangle  \epsilon_j \epsilon_{j'}  \right \rrangle_{i \in \alpha}
     & = w^2 \delta_{jj'}
     +(\epsilon_{\alpha}^2 - w^2) \delta_{ij} \delta_{jj'} .
    \label{eq:<εjεj'>α}
\end{align}
Compared to the prior defined correlator (\ref{eq:<εjεj'>}), we note that the species-resolved version (\ref{eq:<εjεj'>α}) acquires an additional (parenthesized) term, its purpose to ensure that, for $i=j=j'$, the sample is constrained to return $ \llangle \epsilon_{i}^2 \rrangle_{i \in \alpha} = \llangle \epsilon_{\alpha}^2 \rrangle_{i \in \alpha} = \epsilon_{\alpha}^2$.

Next, separating out the local responses/contributions from our disordered Hartree identities
\begin{align}
    \delta n_i 
    & \overset{(\ref{eq:δni})}{=}  M_{ii} \epsilon_i + \sum_{j\neq i}  M_{ij} \epsilon_j,
    \label{eq:δni_2}
    \\
    \phi_i 
    & \overset{(\ref{eq:φi})}{=} \widetilde{M}_{ii}  \epsilon_i + \sum_{j\neq i} \widetilde{M}_{ij} \epsilon_j,
    \label{eq:φi_2}
\end{align}
we compute the species-resolved averages, corresponding to the cross marks shown in the bottom two panels of \Cref{fig:DHstats}, and note that these are finite now
\begin{align}
    \llangle  \delta n_i  \rrangle_{i \in \alpha}
    & =  
    M_{ii}  \llangle \epsilon_i \rrangle_{i \in \alpha}  +  
    \cancelto{0}{  \sum_{j \neq i} M_{ij}  \llangle \epsilon_j  \rrangle _{i \in \alpha}      }
    \, = M_{ii} \epsilon_{\alpha},
    \label{eq:δni_avgα}
    \\
    \llangle \phi_i \rrangle_{i \in \alpha}
    & = 
    \widetilde{M}_{ii} \llangle \epsilon_i \rrangle_{i \in \alpha} +  
    \cancelto{0}{ \sum_{j \neq i} \widetilde{M}_{ij}  \llangle \epsilon_j  \rrangle_{i \in \alpha} }
    = \widetilde{M}_{ii} \epsilon_{\alpha}.
    \label{eq:φi_avgα}
\end{align}
We then compute the mean squares as before
\begin{align}
    \left \llangle    \delta n_i^2  \right \rrangle_{i \in \alpha}     
    &
    = \sum_{jj' }  M_{ij} M_{ij'}  \llangle \epsilon_j \epsilon_{j'} \rrangle_{i \in \alpha}
    \nonumber
    \\
    & = w^2 \sum_{j} M_{ij}^2 + M_{ii}^2 (\epsilon_{\alpha}^2 - w^2) ,
   \label{eq:δni^2_avgα}
    \\
    \left \llangle  \phi_i^2  \right \rrangle_{i \in \alpha} 
    &
    = 
    \sum_{jj' }  \widetilde{M}_{ij}  \widetilde{M}_{ij'}    \llangle     \epsilon_j \epsilon_{j'}  \rrangle_{i \in \alpha}
    \nonumber
    \\
    & = w^2 \sum_{j} \widetilde{M}_{ij}^2 + \widetilde{M}_{ii}^2 (\epsilon_{\alpha}^2 - w^2),
   \label{eq:φi^2_avgα}
\end{align}
though we note that the spreads about the finite intraspecies averages  (\ref{eq:δni_avgα}) and (\ref{eq:φi_avgα}) are now given by the intraspecies variances; these are constructed in the usual manner
\begin{align}
    \left \llangle    ( \delta n_i - \llangle \delta n_i \rrangle_{i\in\alpha} )^2   \right \rrangle_{i\in\alpha}     
    & = 
    \left \llangle    \delta n_i ^2   \right \rrangle_{i\in\alpha} - \llangle    \delta n_i    \rrangle^2_{i\in\alpha},
    \nonumber \\[0.2cm]
    & = w^2 \sum_{j\neq i} M_{ij}^2,
    \label{eq:δni_varα}
    \\[0.2cm]
    \left \llangle  ( \phi_i - \llangle \phi_i \rrangle_{i \in \alpha}  )^2  \right \rrangle_{i\in\alpha}
    & = \left \llangle  \phi_i ^2  \right \rrangle_{i\in\alpha} - \llangle  \phi_i  \rrangle^2_{i\in\alpha}
    \nonumber \\[0.2cm]
    & =  w^2 \sum_{j \neq i} \widetilde{M}_{ij}^2,
    \label{eq:φi_varα}
\end{align}
as is the covariance measuring the co-relationship between $\delta n_i$ and $\phi_i$
\begin{align}
    \llangle \delta n_i \phi_i \rrangle_{i \in \alpha}
    & = \sum_{jj' }  M_{ij} \widetilde{M}_{ij'}  \llangle \epsilon_j \epsilon_{j'} \rrangle_{i \in \alpha}
    \nonumber 
    \\
    & = w^2 \sum_{j} M_{ij} \widetilde{M}_{ij} + M_{ii} \widetilde{M}_{ii} (\epsilon_{\alpha}^2 - w^2),
    \label{eq:δniφi_avgα}
\end{align}
\vspace{-0.4cm}
\begin{align}
    \llangle (\delta n_i - \llangle \delta n_i \rrangle_{i \in \alpha})  (\phi_i - & \llangle \phi_i 
    \rrangle_{i \in \alpha}) \rrangle_{i \in \alpha} 
    \nonumber \\[0.2cm]
    & = \llangle \delta n_i \phi_i  \rrangle_{i \in \alpha} 
    - \llangle \delta n_i \rrangle_{i \in \alpha} \llangle \phi_i  \rrangle_{i \in \alpha}
    \nonumber \\[0.2cm]
    & = w^2 \sum_{j \neq i} M_{ij} \widetilde{M}_{ij} .   
    \label{eq:δniφi_covα}
\end{align}     
Surprisingly, we find that, of the second-order central moments computed above, neither (\ref{eq:δni_varα}) nor (\ref{eq:φi_varα}) nor (\ref{eq:δniφi_covα}) demonstrate any sort of dependence on species $\alpha$ here. This turns out to have rather significant consequences for various physical quantities, as we will now begin to show.

\subsection{Least-squares fit of qV trendlines} \label{subsec:leastsqs}

With the intraspecies statistics obtained in the previous \cref{subsec:DHstats_byspecies}, we may now determine the line of best fit for our species-resolved scatterplots (\Cref{fig:DHstats} bottom row). The optimal fit line parameters for species $\alpha$'s statistics are obtained by minimizing the sample's mean squared error 
\begin{align}
    \mathbb{E}_{\alpha} = \left\llangle \left( \delta n_i - (m_{\alpha} \phi_i + b_{\alpha}) \right)^2 \right\rrangle_{i \in \alpha}
\end{align}
with respect to the species-dependent slope and $y$-intercept, $m_{\alpha}$  and  $b_{\alpha}$, respectively. This provides a pair of equations
\begin{align}
    & \left\{
    \begin{alignedat}{3}
    \frac{ \partial \mathbb{E}_{\alpha} }{ \partial m_{\alpha} } 
    & = \left\llangle 2 \left( \delta n_i - (m_{\alpha} \phi_i + b_{\alpha}) \right)  (- \phi_i) \right\rrangle_{i \in \alpha}
    && & = 0
    \\
    \frac{ \partial \mathbb{E}_{\alpha} }{ \partial b_{\alpha} } 
    & = \left\llangle 2 \left( \delta n_i - (m_{\alpha} \phi_i + b_{\alpha}) \right) (-1) \right\rrangle_{i \in \alpha}
    && & = 0
    \end{alignedat}
    \right.
    \intertext{which, by applying the expectation value's linear properties (additivity, homogeneity), can be rearranged to produce \newline \vspace{-\baselineskip}}
    & \left\{
    \begin{alignedat}{3}
        \vphantom{\frac{ \partial \mathbb{E}_{\alpha} }{ \partial m_{\alpha} }}
         & m_{\alpha} \left\llangle \phi_i^2 \right\rrangle_{i \in \alpha}
         && +
         b_{\alpha} \left\llangle \phi_i \right\rrangle_{i \in \alpha}
         && = \left\llangle \delta n_i \phi_i \right\rrangle_{i \in \alpha},
        \\
        \vphantom{\frac{ \partial \mathbb{E}_{\alpha} }{ \partial b_{\alpha} }}
         & m_{\alpha} \left\llangle \phi_i \right\rrangle_{i \in \alpha}
         && +
         b_{\alpha} \left\llangle 1 \right\rrangle_{i \in \alpha}
         && = \left\llangle \delta n_i \right\rrangle_{i \in \alpha}.
    \end{alignedat}
    \right.
\end{align}
This can be recast as a matrix inversion problem, the solution to which is afforded through standard linear algebra. Thus,
\begin{align}
    m_{\alpha} 
    & = 
    \frac{ 
        \left\llangle 1 \right\rrangle_{i \in \alpha}  \left\llangle \delta n_i \phi_i \right\rrangle_{i \in \alpha}     
        - 
        \left\llangle \delta n_i \right\rrangle_{i \in \alpha}  \left\llangle \phi_i \right\rrangle_{i \in \alpha}       
    }{
        \left\llangle 1 \right\rrangle_{i \in \alpha} \left\llangle \phi_i^2 \right\rrangle_{i \in \alpha}
        - 
        \left\llangle \phi_i \right\rrangle_{i \in \alpha}^2  
    },
    \label{eq:mα_moments}
    \\
    b_{\alpha} 
    & = 
    \frac{ 
         \left\llangle \delta n_i \right\rrangle_{i \in \alpha} \left\llangle \phi_i^2 \right\rrangle_{i \in \alpha}
        - \left\llangle \phi_i \right\rrangle_{i \in \alpha}  \left\llangle \delta n_i \phi_i \right\rrangle_{i \in \alpha}
    }{
        \left\llangle 1 \right\rrangle_{i \in \alpha} \left\llangle \phi_i^2 \right\rrangle_{i \in \alpha}
        - 
        \left\llangle \phi_i \right\rrangle_{i \in \alpha}^2    
    }.
    \label{eq:bα_moments}
\end{align}
Now we can identify here $\phi_i$'s species-resolved variance in the denominator of both $m_{\alpha}$ and $b_{\alpha}$ above, this having been computed previously in (\ref{eq:φi_varα}). Further, we add that the prior's numerator corresponds to the covariance provided in (\ref{eq:δniφi_covα}). Reexpressing these and other remaining quantities using the disordered Hartree moments obtained in the previous \cref{subsec:DHstats_byspecies}, we find that the result reduces to
\begin{align}
    m_{\alpha} 
    & = 
    \frac{
        \sum\limits^{j \neq i} M_{ij} \widetilde{M}_{ij}  
    }{
        \sum\limits_{j \neq i} \widetilde{M}_{ij}^2 
    } ,
    \label{eq:mα_MM~}
    \\
    b_{\alpha} 
    & =
    \frac{ 
        \sum\limits^j \widetilde{M}_{ij} ( M_{ii} \widetilde{M}_{ij} -  M_{ij} \widetilde{M}_{ii} ) 
    }{
        \sum\limits_{j \neq i} \widetilde{M}_{ij}^2    
    } 
    \cdot \epsilon_{\alpha}
    \label{eq:bα_MM~}
\end{align}
Thus, using the linear response language with which we formulated our theory, we can completely describe the qV trendlines which best fit the joint $(\phi_i,\delta n_i)$ intraspecies statistics -- the associated fit parameters primarily involving our charge and Madelung response functions $M$ and $\widetilde{M}$, respectively. Interestingly enough, we observe that the slope of these fit lines $m_{\alpha}$ are entirely independent of chemical species; as alluded to prior, this directly results from the $\alpha$-independence of our second-order central moments.

\newpage
\subsection{Joint statistical line lengths and widths} \label{subsec:widthsAndLengths}

Having obtained the (centers and) spreads of our intraspecies statistics in \cref{subsec:DHstats_overall}, as well as complete parameterizations for their (fitted) qV trendlines in the previous \cref{subsec:leastsqs}, we may now describe just how things are spread both across and along these fits. Elaborating a bit, we note that the species-resolved variances obtained in (\ref{eq:δni_varα}) and (\ref{eq:φi_varα}) provide for us a measure of statistical spread along either the $\delta n_i$ and $\phi_i$ axes, both respectively and independently from one another. What we now wish to determine is: what are the spreads along the major and minor axes of the intraspecies distributions themselves (i.e. in directions parallel and perpendicular to the line of best fit)? 

To acquire these joint statistical line lengths and widths, we first consider an operation which rotates counterclockwise the $(\phi_i,\delta n_i)$ statistics for a given species about its mean such that the resulting $(\phi_i',\delta n_i')$ statistics will have a flat trendline. Taking $(\phi_i,\delta n_i) \rightarrow (x,y)$ to simplify our notation, and doing the same for their primed counterparts, such an operation $\hat{R}$ acts accordingly 
\begin{equation}
    \hat{R} \, \vb{r} = \vb{r}'
    \qquad ; \qquad  
    \vb{r}^{(\prime)}
    =
    \left( \hspace{-0.08cm} \begin{array}{c}
        x^{(\prime)} - \mu_x
        \\
        y^{(\prime)} - \mu_y
    \end{array} \hspace{-0.1cm} \right),
    \label{eq:Rr=r'}
\end{equation}
where $(\mu_x,\mu_y)$ denotes the (intraspecies) average for both the $(\phi_i,\delta n_i)$ and $(\phi_i',\delta n_i')$ datasets, as these means remain unaffected by the pure rotation. Clearly, $\hat{R}$, which acts only to flatten a particular species' trendline, should depend only on the slope of the trendline in question. In fact, we can write
\begin{equation}
    \hat{R}(m) =
    \frac{1}{\sqrt{m + 1}}
    \left( \hspace{-0.1cm} \begin{array}{cc}
        1 & m
        \\
        - m & 1
    \end{array} \hspace{-0.1cm} \right)
    \label{eq:R(m)}
\end{equation}
where $m = m_{\alpha}$ was obtained in (\ref{eq:mα_MM~}) prior, though we drop its $\alpha$ index here to reflect its independence from impurity species. 

With some algebra, we may then use the previous two formulas to express $(x',y')$ in terms of $(x,y)$ -- or equivalently, express $(\phi_i',\delta n_i')$ in terms of $(\phi_i,\delta n_i)$
\begin{align}
    \hspace{-0.1cm}
    \left( \hspace{-0.08cm} \begin{array}{cc}
        x'
        \\
        y'
    \end{array} \hspace{-0.1cm} \right)    
    =
    \frac{1}{\sqrt{m^2 + 1}} 
    \left( \hspace{-0.08cm} \begin{array}{cc}
        x - \mu_x(1-\sqrt{m^2 + 1}) + m(y-\mu_y)
        \\
         y - \mu_y(1-\sqrt{m^2 + 1}) - m(x-\mu_x)
    \end{array} \hspace{-0.1cm} \right) \! \!. 
    \hspace{-0.1cm}
\end{align}
Finally, we compute the (intraspecies) statistical moments for these rotated (primed) variables, as we did previously in \cref{subsec:DHstats_byspecies} for their original (unprimed) counterparts. Following several more steps of algebra and invocations of the unprimed moment identities produced in this prior \cref{subsec:DHstats_byspecies}, we obtain the following result
\begin{align}
    \sigma_{x'}^2 & = \sigma_x^2 + \frac{ m^2 }{m^2 + 1} \left(  \sigma_x^2  + \sigma_y^2 \right),
    \label{eq:σx'sqd}
    \\
    \sigma_{y'}^2 & = \frac{1}{m^2 + 1} \left( \sigma_y^2 - m^2 \sigma_x^2 \right),
    \label{eq:σy'sqd}
\end{align}
where $(\sigma_{x}^2,\sigma_y^2)$ represents the (intraspecies) variances (\ref{eq:φi_varα},\ref{eq:δni_varα}) along the $(\phi_i, \delta n_i)$ axes, while $(\sigma_{x'}^2,\sigma_{y'}^2)$ specifies the variance in the rotated $(\phi_i',\delta n_i')$ frame, and hence the statistical spread (along, about) the line of best fit.  Note that these both remain $\alpha$-independent, as $m,\sigma_x^2$, and $\sigma_y^2$ have all been shown to be.

\clearpage

\onecolumngrid

\begin{figure}[hbt!]
    \centering
    \begin{tabular}{c c}
        \includegraphics[trim={0 0.3cm 0 1.5cm}, clip, width=0.475\linewidth]{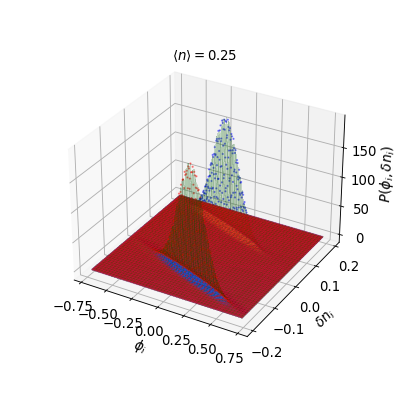}
        &
        \includegraphics[trim={0 0.3cm 0 1.5cm}, clip, width=0.475\linewidth]{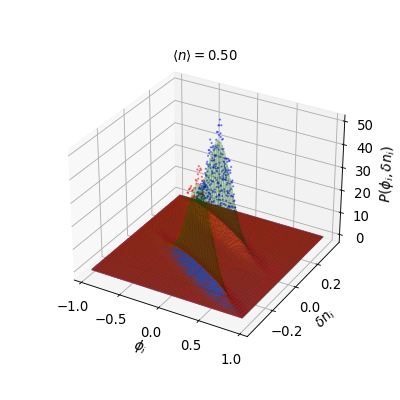}
    \end{tabular}
    \caption{Joint probability distributions $P(\phi_i, \delta n_i)$ for the local intraspecies statistics of the (left) quarter-filled and (right) half-filled cubic models with equiprobable binary disorder. The data points shown here are colored red/blue corresponding to the positive/negative impurity species, and these represent both the histogramming and subsequent normalization of the scatterplots contained in the bottom row of \Cref{fig:DHstats}. Our two-dimensional histograms span $(100\times100)$-grids each, and are compared against the bivariate normal distributions (green surface plots) whose arguments are given by the disordered Hartree moments derived in \cref{subsec:DHstats_byspecies}. } 
    \label{fig:bivarNorm}
\end{figure}

\twocolumngrid

\subsection{Bivariate normal approximation for the joint probability distributions}

For a more complete description of the joint statistical relationship between $\delta n_i$ and $\phi_i$, we revisit the scatterplots presented in \Cref{fig:DHstats}'s bottom row and consider the probability distribution which produces them. First, we observe that the scatterplotted data points all do combine to form fairly ellipsoidal contours. Furthermore, we recall that the individual statistics of either $\delta n_i$ and $\phi_i$ did present with some Gaussian qualities, these being quite apparent at quarter-filling, though perhaps more envelopic at half-filling. Together, these observations suggest that the joint probability distribution $P(\phi_i, \delta n_i)$ may obey, or otherwise be decently approximated, by a two-dimensional generalization of the usual Gaussian curve.  

We verify this last statement in \Cref{fig:bivarNorm} above where we again use the earlier example of equiprobable binary disorder for both the quarter- and half-filled cubic models. Upon histogramming and normalizing the (intraspecies) scatterplots presented in \Cref{fig:DHstats}'s bottom row, we obtain the joint probability distributions $P(\phi_i,\delta n_i)$ for either positive or negative binary species (\Cref{fig:bivarNorm} displaying bin heights as data points above). We then compare these statistics against corresponding bivariate normal distributions (green surface plots) which most generally assume the form given by (\ref{eq:Pxy}) below.

\onecolumngrid

\begin{equation}
    P(x, y) = - \frac{1}{2\pi \sigma_x \sigma_y \sqrt{1-\rho^2}} \,
    \exp[ \frac{1}{2(1-\rho^2)} \left(
        \frac{(x-\mu_x)^2}{\sigma_x^2} 
        + 2 \rho \, \frac{(x-\mu_x)}{\sigma_x}\frac{(y-\mu_y)}{\sigma_y} 
        + \frac{(y-\mu_y)^2}{\sigma_y^2} 
    \right) ]
    \qquad ; \qquad 
    \rho = \frac{ \sigma_{xy} }{ \sigma_x \sigma_y }.
    \label{eq:Pxy}
\end{equation}    

\twocolumngrid

Note that this standard distribution is parameterized entirely by moments only up to and including the second-order%
\footnote{This in a manner similar to that of its univariate counterpart -- i.e. the Gaussian/normal distribution, which requires only the first (average) and second (variance) moments for complete specification.}, 
all of these being provided by our theory and computed prior in \cref{subsec:DHstats_byspecies}. Note also that, as was done in the previous \cref{subsec:widthsAndLengths}, we simplify our notation in (\ref{eq:Pxy}) by taking $(x,y)\rightarrow(\delta n_i, \phi_i)$ and using $\mu_{x,y}$ and $\sigma_{x,y}^2$ to denote the (intraspecies) averages and variances, respectively, of $x,y$. We further assign $\sigma_{xy}$ as their covariance. This shorthand we describe are all captured by the following map (\ref{eq:DHmoms_xymap}). 

\begin{equation}
    \left\{
    \begin{aligned}
        \mu_{x} & \qquad \longrightarrow \qquad \widetilde{M}_{ii} \epsilon_{\alpha} 
        \\[0.3cm]
        \mu_{y} & \qquad \longrightarrow \qquad  M_{ii} \epsilon_{\alpha} 
        \\[0.3cm]
        \sigma_{x}^2 & \qquad \longrightarrow \qquad  w^2 \sum_{j \neq i} \widetilde{M}_{ij}^2 
        \\
        \sigma_{y}^2  & \qquad \longrightarrow \qquad w^2 \sum_{j \neq i} M_{ij}^2
        \\
        \sigma_{xy}  & \qquad \longrightarrow \qquad w^2 \sum_{j \neq i} M_{ij} \widetilde{M}_{ij} 
    \end{aligned}
    \right.
    \label{eq:DHmoms_xymap}
\end{equation}

\clearpage

Now by comparing our disordered Hartree statistics to the assumed distribution in \Cref{fig:bivarNorm}, we find excellent agreement particularly in the quarter-filled example which, as was previously noted, demonstrates sufficiently Gaussian statistics thanks to the weaker screening efficiency at smaller carrier density. In the half-filled case, where stronger screening results in more nontrivial statistical features, we still find that the bivariate normal distribution (\ref{eq:Pxy}) can provide somewhat of a decent approximation for the joint probability; although (\ref{eq:Pxy}) certainly is not designed to capture the sharper horns and peaks discernible in the half-filled statistics, we find that it provides at least a satisfactory description of the Gaussian envelope overarching these finer details.

\subsection{Isoelectronic doping with binary disorder} \label{subsec:scanConc}

Thus far, the recurring example of equiprobable binary disorder has allowed us to develop a simpler understanding of the disorder-driven charge transfer and Madelung field fluctuations within our first-order disordered Hartree framework. Now to gain further insight into the problem, we shall study the effects that varying disorder will have by tuning through the relative impurity concentrations between the binary species.

Before we proceed with this though, we note that, if we wish to apply the knowledge and identities we have carefully developed since the start of the current \cref{sec:DHstats}, then we must account for an additional 
subtlety. See, when we had first introduced this equiprobable binary problem, we had formally assigned either species $\alpha \in \{ + , -  \} $ to each $(j\in N)^{\textrm{th}}$ lattice site, each associated with a site-energy $\epsilon_{\pm} = \pm w$ and having equal probability of occurrence $p_{\pm} = 1/2$. If, however, we now wish to vary the relative concentrations by tuning through $p_\pm$ while naively fixing $\epsilon_{\pm}$, what we find is that
\begin{align}
    \llangle \epsilon_j \rrangle 
    & = p_+ \epsilon_+ - p_- \epsilon_- = w \, (p_+ - p_-) \neq 0 
\end{align}
as $ p_+ \neq p_-$ away from equal probabilities%
\footnote{Note that by probability conservation/normalization, $\sum_{\alpha} p_{\alpha} = 1$. Hence $p_{\pm} = 1-p_{\mp}$ in the binary case using our current language.}.%
Thus we fail to meet the previous criteria for conserving statistics (\ref{eq:<εj>}), and all of the results which follow become no longer valid. 

Now although we note that it is possible to redevelop our statistical formalism to account for such finite $\llangle \epsilon_j \rrangle$s, the remaining issue with this method is that it still runs the risk of entering a pathology, because with fixed $\epsilon_{\pm}$, $\llangle \epsilon_j \rrangle$ may then vary freely between $\epsilon_{\pm}$ at different concentrations. And this, for us, results in a linearly perturbed solution which may be drastically dissimilar -- i.e. in average site-energy and hence number of electrons -- from that which describes the bare/reference system $\hat{H}_0$; this is in blatant conflict with the assumptions underlying our perturbative approach. 

Thus, for a more satisfactory method of varying the binary impurity concentrations, we employ the following procedure:
\begin{enumerate}[nosep,leftmargin=*]
    \item we first relax the prior constraint  ($\epsilon_{\pm} = \pm w$) on site-energies, these now becoming concentration-dependent.
    \item as we vary concentrations ($p = p_+$ associated with $\epsilon_+$ and $1-p = 1-p_+ = p_-$ associated with $\epsilon_-$), we recompute $\epsilon_+$ and $\epsilon_-$ at each step such that the statistical criteria (\ref{eq:<εj>}) is satisfied. Consequently,
    \begin{equation}
        \left.
        \begin{aligned}
            \llangle \epsilon_j \rrangle 
            & = p \, \epsilon_+ + (1-p) \, \epsilon_- 
            \\
            & = 0
        \end{aligned}
        \right\}    
        \quad \longrightarrow \quad 
        \epsilon_- = - \frac{p \, \epsilon_+ }{ 1 - p }.
        \label{eq:ε-}
    \end{equation}
    \item simultaneously, we fix the difference between site-energies $\Delta \epsilon = \epsilon_+ - \epsilon_- $ across all $p$. This then implies that
    \begin{equation}
        \left.
        \begin{aligned}
            \Delta \epsilon 
            & = \epsilon_+ - \epsilon_- 
            \\
            & \overset{\mathclap{(\ref{eq:ε-})}}{=}  \epsilon_+ \left( \frac{1}{1-p} \right) 
        \end{aligned}
        \right\}
        \quad \longrightarrow \quad
        \epsilon_+ = \Delta \epsilon (1-p).
    \label{eq:ε+}
    \end{equation}
\end{enumerate}
Through this procedure, we constrain our statistics to the desired form while preserving the average electron density across all $p$. This is done in a manner which precludes any possible ambiguity associated with impurity degrees of freedom when varying concentrations -- all of this burden placed solely on both $p$ and $\Delta \epsilon$. With these two quantities, we may completely characterize the disorder of the binary problem since, through (\ref{eq:<εjεj'>}), we have 
\begin{align}
    \left.
    \begin{aligned}
        \llangle \epsilon_j^2 \rrangle & = w^2
        \\
        & = p \, \epsilon_+^2 + (1-p) \epsilon_-^2
        \\
        & 
        \overset{\mathclap{ (\ref{eq:ε-}) } }{=} 
        p \, \epsilon_+^2 \left( \frac{1}{1-p}  \right) 
        \\
        & 
        \overset{\mathclap{ (\ref{eq:ε+}) } }{=} 
        p  (1-p) \Delta \epsilon^2  
    \end{aligned}
    \right \}
    \, \, \longrightarrow \, \, \, \, \,
    w^2 & = p (1-p) \Delta \epsilon^2,
    \label{eq:w^2}
\end{align}
where we note that, when $p=\{0,1\}$, we then have $w = 0$ as this brings us to the clean/uniform limit that is free from chemical disorder. Thus, to reiterate, as we scan through concentration $p$, we can hold the binary site-energy difference $\Delta \epsilon$ constant which then constrains all remaining quantities associated with impurity disorder. 

\subsection{Concentration- and filling-dependence of statistical features}

Following the above-outlined procedure, we collect an assortment of quantities that describe the qV trendlines and their underlying statistics. We do so for various (positive) impurity concentrations $p$ as well as fillings $\langle n \rangle$ when $\Delta \epsilon = 2$ is fixed. We include our results in  \Cref{fig:concTrends} below, where the data sets at each $\langle n \rangle$ are further rescaled by their half-concentration values (or their absolute values). This collapses the curves and reveals their universal $p$-dependence. For consistency, we continue to treat the same cubic model used throughout our work. We further note that, by considering binary disorder where $\Delta \epsilon = 2$ is held fixed, our data survey therefore includes the same equiprobable binary profile ($\epsilon_{\pm} = \pm 1$) simulated previously in \Cref{fig:DHstats} when concentration $p = 0.5$. 
\clearpage 

\onecolumngrid

\begin{figure}[hbt!]
    \centering
    \includegraphics[trim={0.3cm 0.2cm 0 0}, clip, width=0.8\linewidth]{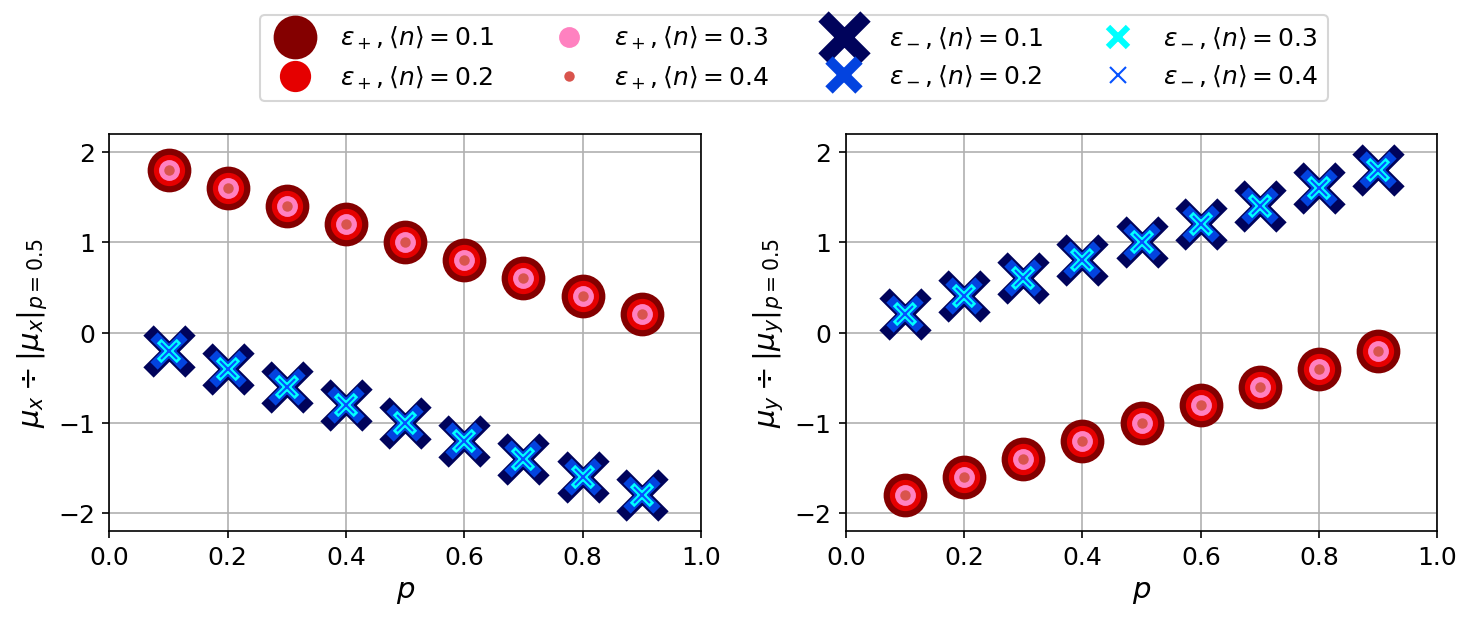}
    \\
    \includegraphics[trim={0.3cm 0.2cm 0 0}, clip, width=0.4\linewidth]{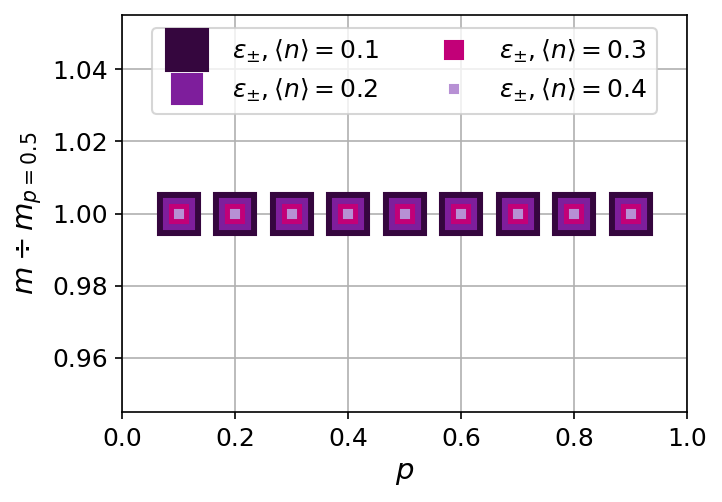}
    \\
    \includegraphics[trim={0.3cm 0.2cm 0 0}, clip, width=0.8\linewidth]{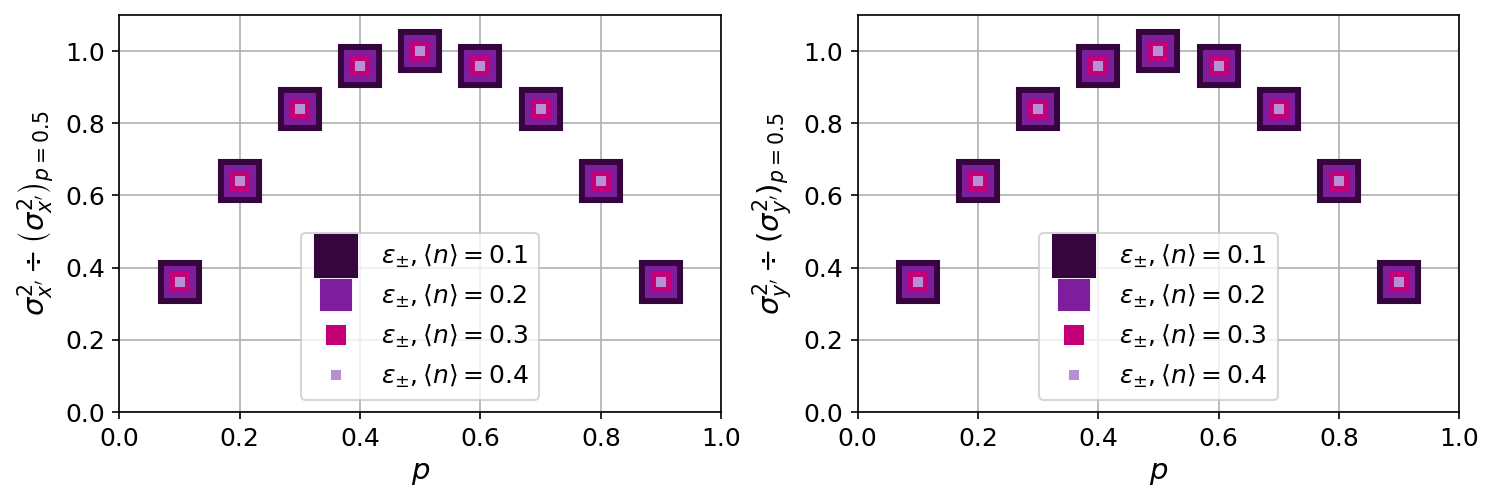}
    \caption{Disordered Hartree statistical averages (top row), slopes (middle row), and line dimensions (bottom row) characterizing the qV trendlines of the binary disorder problem with varying concentration $p$. These have been rescaled by their half-concentration values to reveal their universal $p$-dependence irrespective of filling $\langle n \rangle$.}
    \label{fig:concTrends}
\end{figure}

\twocolumngrid

From \Cref{fig:concTrends} above, we observe the following:
\begin{enumerate}[nosep,leftmargin=*]
    \item the intraspecies averages, $\mu_{x}$ and $\mu_y$, both display a linear dependence in $p$. This can be understood by recalling from (\ref{eq:DHmoms_xymap}) that $\mu_{x,y} \sim \epsilon_{\alpha}$, where $\alpha \in \{+,-\}$ in the current binary example. Further noting from both (\ref{eq:ε-}) and (\ref{eq:ε+}) prior that $\epsilon_{\pm} \sim \mp p $ when $\Delta \epsilon$ is fixed, the source of the linear dependence becomes clear. 
    \item the species-independent slopes $m$ of the qV trendlines shows no dependence on concentration. As derived in (\ref{eq:mα_MM~}), this quantity depends only on the response functions $M$ and $\widetilde{M}$ themselves without any regard for the disorder profile. This explains the lack of explicit $p$-dependence.
    \item the species-independent qV line lengths $\sigma_{x'}^2$ and line widths $\sigma_{y'}^2$ both demonstrate quadratic dependence in $p$. This we understand by recognizing that, with $m $ being $p$-independent, (\ref{eq:σx'sqd}) and (\ref{eq:σy'sqd}) prescribe both of these quantities as some linear combination of $\sigma_x^2$ and $\sigma_y^2$. Next, using both (\ref{eq:DHmoms_xymap}) and \ref{eq:w^2}, we deduce that $\sigma_{x,y}^2 \sim w^2 \sim p(1-p)$. It therefore stands to reason that $\sigma_{x',y'}^2$ would be quadratic in $p$ as well.
\end{enumerate}

\begin{figure}[hbt!]
    \centering
    \includegraphics[trim={0.3cm 0.2cm 0 0}, clip, width=0.8\linewidth]{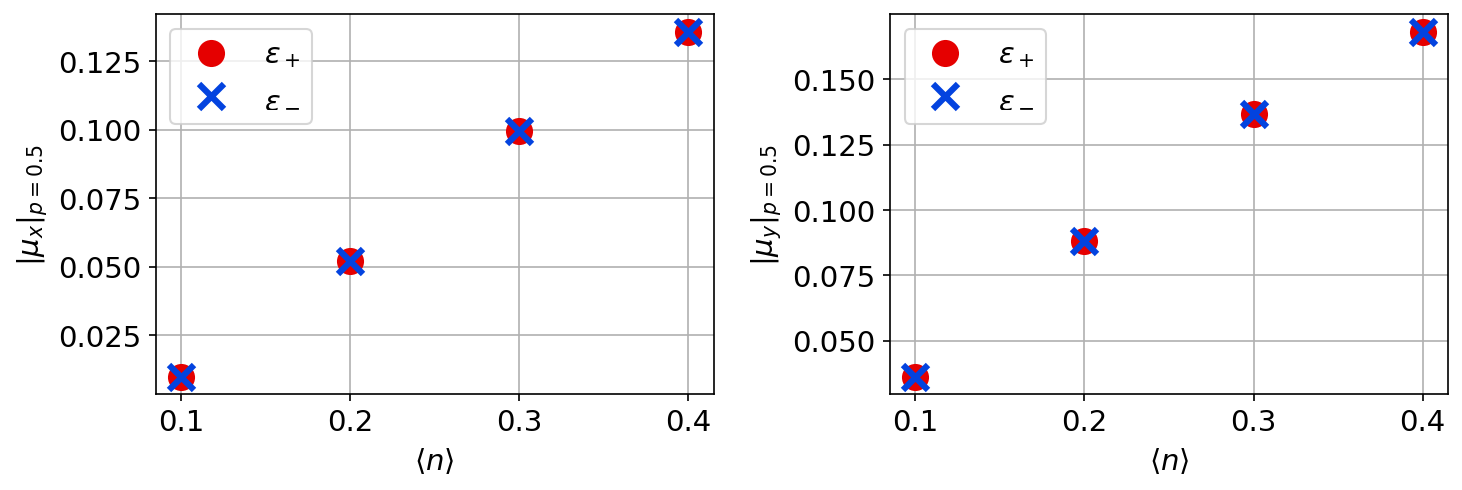}
    \\
    \includegraphics[trim={0.3cm 0.2cm 0 0}, clip, width=0.4\linewidth]{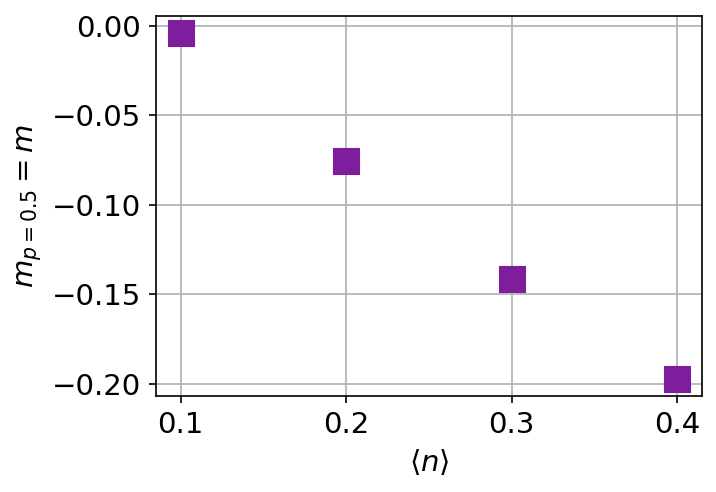}
    \\
    \includegraphics[trim={0.3cm 0.2cm 0 0}, clip, width=0.8\linewidth]{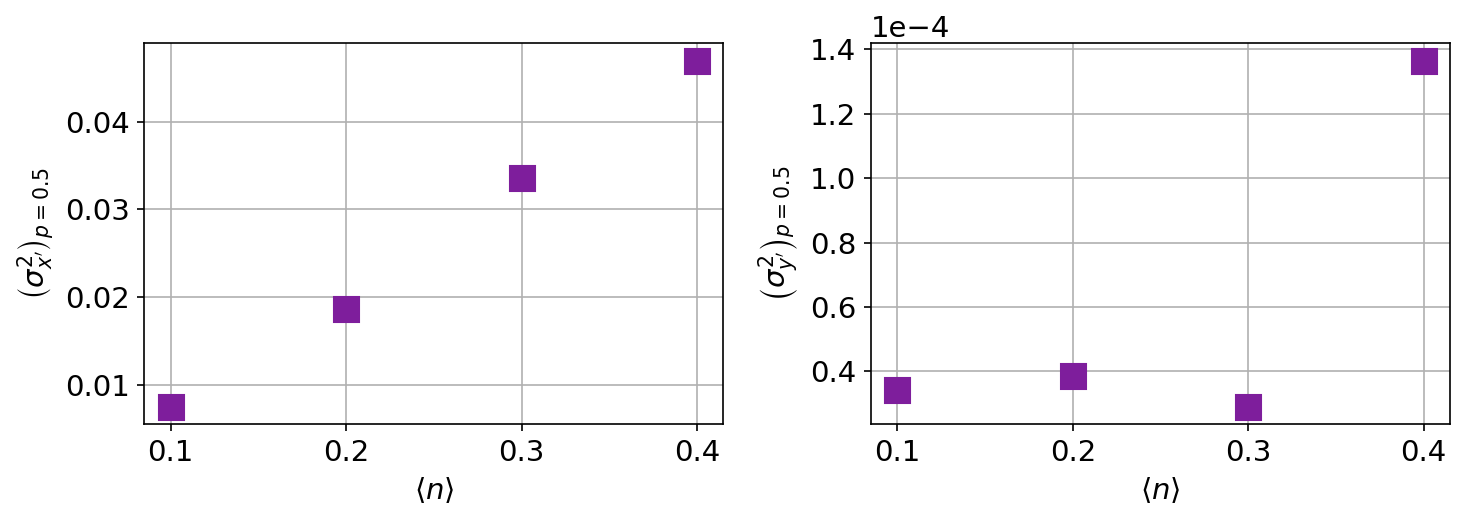}
    \caption{Disordered Hartree statistical averages (top row), slopes (middle row), and line dimensions (bottom row) characterizing the qV trendlines in the binary disorder problem with fixed concentration $p=0.5$. These half-concentration values correspond to the rescaling factors that were used in \Cref{fig:concTrends} to reveal these quantities' universal $p$-dependence.}
    \label{fig:rescaleFactors}
\end{figure}

Now lastly, as mentioned, rescaling each data set by their half-concentration values allows us to collapse the various curves which differ by filling $\langle n \rangle$ onto one another and deduce their universal $p$-dependence. We note now that these rescaling factors alone provide us with $\langle n\rangle$-dependent statistical trends which have been replotted in \Cref{fig:rescaleFactors} below to illustrate how different quantities depend on filling or carrier concentration. From these, we find that the disordered Hartree averages, $\mu_x$ and $\mu_y$, both grow in magnitude as $\langle n \rangle$ is increased. This implies a larger interspecies statistical splitting as carrier concentration grows. Additionally, we find that the qV trendline slopes $m$ tend to flatten out, while the statistical line lengths $\sigma_{x'}^2$ of the qV trendline will increase, both with higher $\langle n \rangle$. Alternatively, we note that there appears to be no simple dependence of the line widths $\sigma_{y'}^2$ on filling.

\onecolumngrid

\twocolumngrid


\section{CONCLUSIONS}

We have developed here a theory of chemically disordered alloys by constructing a minimal-working one-band model of interacting fermions within a random environment. Our work reveals the key physical processes which contribute to the electronic charge transfer and Madelung field fluctuations inherent in these materials; such processes include the Friedel oscillations and the screening of electrostatic potentials generated by each chemical defect. When many such defects are present, as is the case in these random alloys, we find that all of their associated effects then interfere to produce the fluctuating charge and Madelung field profiles found throughout the bulk. 

Within our simplified framework, we have also analyzed the statistics associated with these features and have developed a formal way of describing them. Such statistics are all encompassed by the linear joint statistical qV relationships shared between the charge transfer and Madelung fields. These were prior identified in more comprehensive first-principles studies of both conventional and high-entropy alloys, yet have remained poorly understood until now. Using our simplified model, which we have shown to properly recover these details, we have been able to answer several of the most basic yet pressing questions behind them. This includes what governs their various statistical features and how these all depend on different physical quantities, including impurity concentration as well as carrier density. Thus, we provide here the answers to several important questions concerning the internal mechanism of generic alloys, which lend themselves towards furthering developments in the new field of high-entropy alloys.

Lastly, we add that, by providing a complete statistical description of these qV trendlines, our work also opens up the avenue for providing systematic corrections to modern methods of disorder modeling which currently fail to account for their associated physics. This includes the conventional CPA approach, improvements to which may be further developed in future work.

\begin{acknowledgements}
Work in Florida (WGDH and VD) was supported by the NSF Grant No. 1822258, and the National
High Magnetic Field Laboratory through the NSF Cooperative Agreement No. 1644779 and the State of Florida. HT was supported by NSF OAC-1931367 and NSF DMR-1944974 grants. KMT was partially supported by NSF DMR-1728457 and NSF OAC-1931445. YW was partially supported by NSF OAC-1931525. 
\end{acknowledgements}

\appendix

\section{FRIEDEL OSCILLATIONS IN THE DISORDERED HARTREE FRAMEWORK} \label{app:FriedelOsc}

To recover the Friedel oscillations in our linearized disordered Hartree framework, we consider the standard model of an ideal Fermi gas and embed within it just a single impurity defect. In our perturbative language, this just means that the band-dispersion of our bare reference system takes the usual quadratic form, while the perturbation is caused by a lone impurity potential. No interactions are present in this picture, and, placing the impurity of strength $\epsilon_j$ on site $j$, the full Hamiltonian is then given by (compare with (\ref{eq:H})) 
\begin{equation}
    \hat{H} = \hat{H}_0 + \hat{V}
    \qquad ; \qquad 
    \left\{
    \begin{aligned}
        \\[-0.35cm]
        \hat{H}_0 
        & = \sum_{\vb{k} } \frac{1}{2}|\vb{k}|^2 \,  \hat{c}^{\dagger}_{\vb{k}}\hat{c}_{\vb{k}}
        \\
        \hat{V} 
        & = \sum_i \epsilon_j \, \delta_{ij} \, \hat{c}^{\dagger}_i \hat{c}_i 
    \end{aligned} .
    \right.
    \label{eq:H_FriedelApp}
\end{equation}
Of course, in the standard textbook approach, we may often benefit from taking additional idealizations, though we will postpone these until they become necessary to achieve the expected analytical result. 

For now, we exploit the fact that the above has the same form as our original model (\ref{eq:H}), where comparing with this, we simply replace $\xi_{\vb{k}} \rightarrow |\vb{k}|^2/2$ and $\widetilde{\epsilon}_{i} \rightarrow \epsilon_j \, \delta_{ij}$. Thus, the formalism we had developed in \cref{subsec:DHmodel,subsec:PTgen,subsec:linSC} may still apply, and we can then immediately write down
\begin{align}
        \delta n_i = M_{ij} \epsilon_j \qquad \qquad ; \qquad \qquad M_{ij} = A \delta_{ij} + B_{ij},
        \label{eq:δni_FriedelApp}
\end{align}
as we had in equation (\ref{eq:deltani_singleimp}), to treat this non-interacting, single-impurity problem. The (first-order) density response function $M_{ij}$ is given now in this non-interacting limit by our perturbative expansion coefficients, or alternatively, the Lindhard function, these being defined in (\ref{eq:coefficientAquantum}) and (\ref{eq:coefficientBijquantum}) of the main text. Thus, to compute
\begin{equation}
    M_{ij} = -\frac{1}{\pi} \int_{-\infty}^{\mu}\textrm{Im} [{G_0}^2_{ij} ] \, \dd \omega 
    \label{eq:Mij_FriedelApp}
\end{equation}
we must first obtain the bare Green's function matrix elements in the position-space representation. 

From (\ref{eq:H_FriedelApp}), we know $\hat{H}_0$ is diagonal in $\vb{k}$-space. And with 
\begin{align}
    \hat{G}_{0} 
    & \overset{(\ref{eq:GF})}{=} (\omega^+ - \hat{H}_{0} )^{-1}
    \qquad ; \qquad
    \omega^+ = \lim_{\eta \rightarrow 0^+} (\omega + i\eta),
    \label{eq:G0_FriedelApp}
\end{align}
we similarly obtain the Green's function's diagonal entries in the momentum eigenspace
\begin{align}
    {G_0}_{\vb{k}}
    = (\omega^+ - \hat{H}_{0} )^{-1}_{\vb{k}}
    = \frac{1}{\omega^+ - \frac{1}{2}|\vb{k}|^2}.
\end{align}
We will now proceed to Fourier transforming this back over to a position-space representation. 

Note that, in practice, such transformations implicitly involve finite sums over discrete $\vb{k}$-grids which span the first Brillouin zone. However, as alluded to prior, we now facilitate analytical/integration methods by taking the thermodynamic limit where $\vb{k}$ becomes a continuous variable. Moreover, we treat real-space as continuous, moving on from the lattice model that we have developed and used outside of this appendix; this further allows for $\vb{k}$ to span over all $\mathbb{R}^3$. Thus our lattice transform becomes a continuous one following the appropriate modifications shown below, where we are now explicit with summation/integration bounds. We add further that, to better accomodate ${G_0}_{\vb{k}}$'s spherical symmetry, we use spherical coordinates in our Fourier transform.
\begin{align}
    {G_0}_{ij}  
    &  = \frac{1}{N} \sum_{\vb{k} \in \textrm{BZ} }  \frac{1}{\omega^+  - \frac{1}{2} |\vb{k}|^2  } \; e^{i \vb{k} \bigcdot (\vb{r}_i - \vb{r}_j)}
    \\
    & \, \, \Big\downarrow - \textrm{ \scriptsize  continuum, thermodynamic limit}
    \nonumber
    \\[0.1cm]
    {G_0}_{ij}  
    & = \frac{ \Omega_0 }{ (2\pi)^3 } \int_{\mathbb{R}^3}  \frac{1}{\omega^+ - \frac{1}{2} |\vb{k}|^2 } \; e^{i \vb{k}\bigcdot (\vb{r}_i-\vb{r}_j ) } \, \dd^3\vb{k}
    \nonumber \\
    & = \frac{ \Omega_0 }{ (2\pi)^3} \int_0^{\infty} \!\! \int_0^{\pi} \!\! \int_0^{2\pi}  \frac{e^{i k |\vb{r}_i-\vb{r}_j| \cos{\theta} }}{\omega^+ - \frac{1}{2} k^2 } \, k^2 \sin{\theta}  \, \dd k \, \dd\theta \, \dd\phi
    \nonumber \\ 
    \hspace{-1.5cm}  & = \frac{ \Omega_0}{(2\pi)^2} \int_0^{\infty} \frac{k^2}{\omega^+ - \frac{1}{2} k^2} \, \dd k \int_0^{\pi} \sin{\theta}  \, e^{ik|\vb{r}_i-\vb{r}_j|\cos{\theta}} \dd \theta    \hspace{1.5cm} 
    \nonumber  \\ 
    & =  \frac{\Omega_0}{2\pi^2} \frac{1}{|\vb{r}_i-\vb{r}_j|}  \int_0^{\infty} \frac{k \sin{(k|\vb{r}_i-\vb{r}_j|})}{\omega^+ - \frac{1}{2} k^2}   \, \dd k.
    \intertext{Between the final three lines above, we integrate out the azimuthal and polar angles, $\phi$ and $\theta$, respectively -- the prior being trivial and the latter manageable by standard $u$-substitution. Expressing this last result as}
    \hspace{-2cm} 
    {G_0}_{ij}  
    & =\frac{\Omega_0}{2\pi^2} \frac{1}{|\vb{r}_i-\vb{r}_j|} \frac{d}{d|\vb{r}_i-\vb{r}_j|}  \left[  \int_0^{\infty}  \frac{\cos{(k|\vb{r}_i-\vb{r}_j|)}}{\frac{1}{2}k^2 - \omega^+} \, \dd k \right], 
    \hspace{-2cm} 
\end{align}
we next note that the square-bracketed integral can be evaluated by first restoring $\omega^+$ from (\ref{eq:G0_FriedelApp}) and relabeling its real part using an auxiliary momentum variable $k_{\omega} = \sqrt{2\omega}$. This leaves the task of contour integration to achieve the final form of the (bare) free particle Green's function. Through the residue theorem, we essentially find  
\begin{align}
    {G_0}_{ij}  
    & = \frac{\Omega_0}{\pi^2} \frac{1}{|\vb{r}_i-\vb{r}_j|} \frac{d}{d|\vb{r}_i-\vb{r}_j|}  \left[ \lim_{\eta \rightarrow 0^+} \int_0^{\infty}  \frac{\cos{(k|\vb{r}_i-\vb{r}_j|)}}{k^2 - k_{\omega}^2+ i\eta  } \dd k \right],
    \nonumber
    \\
    & = \frac{\Omega_0}{\pi^2} \frac{1}{|\vb{r}_i-\vb{r}_j|} \frac{d}{d|\vb{r}_i-\vb{r}_j|}  \left[ \frac{i \pi}{2} \frac{ e^{i k_{\omega}|\vb{r}_{i} - \vb{r}_{j}| }}{ k_{\omega}} \right],
\end{align}
or more simply,
\begin{equation}
        {G_0}_{ij} = - \frac{\Omega_0}{2\pi} \frac{1}{|\vb{r}_i - \vb{r}_j|}   e^{i k_{\omega} |\vb{r}_i - \vb{r}_j|}.  
\end{equation}

Now with this free particle Green's function, we can finally compute the (first-order) density response function
\begin{align}
    M_{ij} 
    & \overset{(\ref{eq:Mij_FriedelApp})}{=} -\frac{1}{\pi} \int_{-\infty}^{\mu}\textrm{Im} [{G_0}^2_{ij} ] \, \dd \omega 
    \nonumber
    \\
    & \overset{\hphantom{(\ref{eq:Mij_FriedelApp}) } }{=} - \frac{\Omega_0}{4\pi^3} \frac{1}{|\vb{r}_i - \vb{r}_j|^2}  \int_{-\infty}^{\mu} \textrm{Im} [e^{2 i k_{\omega} |\vb{r}_i - \vb{r}_j|} ] \, \dd \omega ,
\end{align}
to recover the Friedel oscillations. Note however that, for $\omega < 0$, $k_{\omega} = \sqrt{2\omega}$ becomes imaginary, which results in either unphysical or exponentially suppressed contributions to the result. Indeed the bare band dispersion $|\vb{k}|^2/2$ is positive-definite, which obviates any need to consider negative $\omega$s. We thus take the opportunity to modify the lower cutoff of our integral such that 
\begin{align}
    M_{ij} 
    & =  - \frac{\Omega_0}{4\pi^3} \frac{1}{|\vb{r}_i - \vb{r}_j|^2}  \int_{0}^{\mu} \sin[2 k_{\omega} |\vb{r}_i - \vb{r}_j| ] \, \dd \omega
    \\
    & =  - \frac{\Omega_0}{4\pi^3} \frac{1}{|\vb{r}_i - \vb{r}_j|^2}  \int_{0}^{k_F} \sin[2 k_{\omega} |\vb{r}_i - \vb{r}_j| ] \, k_{\omega} \,  \dd k_{\omega} 
\end{align}
where we change integration variables by exchanging $\omega=k_{\omega}^2/2$ for $k_{\omega}$, and we further introduce the Fermi wavevector $k_F = \sqrt{2\mu}$ as the new upper cutoff. This last line is straightforward to evaluate using integration-by-parts, the result being
\begin{align}
    \hspace{-0.5cm}
    M_{ij}   
     =  \frac{\Omega_0 k_F}{(2\pi)^3}    
    \left[ \;  \frac{ \cos{(2k_F |\vb{r}_i - \vb{r}_j| ) } }{ |\vb{r}_i - \vb{r}_j|^3 } - \frac{ \sin{ (2k_{F} |\vb{r}_i - \vb{r}_j| ) }} {2k_F |\vb{r}_i - \vb{r}_j|^4} 
    \right].
    \hspace{-0.5cm}
\end{align}
This furnishes for us the Friedel oscillations which are generated around a single impurity defect in the ideal Fermi gas. To leading-order in impurity displacement $|\vb{r}_i - \vb{r}_j|$, we recover the standard result where $M_{ij}$ and, through (\ref{eq:δni_FriedelApp}), $\delta n_i$ both oscillate with twice the Fermi wavevector and decay as an inverse cube.

\section{CLASSICAL DISORDERED HARTREE THEORY} \label{app:classicalDH}

We construct here the classical, high-temperature analogue of the disordered Hartree model, as well as its perturbative solution relating fluctuating charge transfer and Madelung field distributions in chemically disordered metals. What follows shall correspond to the theory we had developed for the quantum (zero-temperature) limit in the main body of this work, particularly in \cref{sec:theory}.

We start with the quantum Hamiltonian we had originally presented in (\ref{eq:H}), and suppress now the hopping/tunneling term ($\{ t,\hat{H}_0 \} \rightarrow 0$), as the associated quantum fluctuations become washed out at higher temperatures. Replacing also each quantum operator (hatted: $\hat{ }$ ) with their classical counterparts (checked: $\check{ }$ ), what remains is a purely local model that is diagonal in the site basis
\begin{align}
    \check{H} 
    = \check{V}
     = \sum_i \widetilde{\epsilon}_i \,  \check{n}_i
    \qquad ; \qquad 
    \widetilde{\epsilon}_i = \epsilon_i + \phi_i.
    \label{eq:classicalH}
\end{align}
Note here that the quantum number operator $(\hat{n}_i = \hat{c}^{\dagger}_i \hat{c}_i) $  is exchanged for its classical complement, the latter only assuming integer occupation values subject to Pauli exclusion  $(\check{n}_i = 0,1)$. Now by adding in an explicit chemical potential term, we can construct the grand canonical Hamiltonian
\begin{align}
    \check{K} 
    & = \check{H} -\mu \sum_{i} \check{n}_i  = \sum_{i} h_i \, \check{n}_i
    \qquad  ;  \qquad
    h_i = \widetilde{\epsilon}_i - \mu,
    \label{eq:classicalK}
\end{align}
the eigenvalues ($h_i$) of which appear in various relevant and useful thermodynamic quantities. 

Of particular and immediate interest to us is the local charge distribution. And in the current high-temperature regime, these are governed by thermal fluctuations in $\check{n}_i$ which are well-described by classical statistical mechanics. Thus, with local site indices ($i$) offering a good basis to work in, we can immediately determine that the average on-site occupancy is captured by the Fermi-Dirac distribution
\begin{align}
    n_i = \langle \check{n}_i \rangle = n_F(h_i) 
    = \frac{1}{e^{\beta h_i} + 1}
    \overset{ (\textrm{\ref{eq:classicalK}}) }{=} \frac{1}{e^{\beta (\widetilde{\epsilon}_i - \mu ) } + 1},
    \label{eq:niFermi}
\end{align}
where $\beta = T^{-1}$ is the inverse temperature, and the angled braces around the classical number operator $\hat{n}_i = 0,1$ are understood to represent its classical/thermal average.

Now in our perturbative language, this above is analogous to the "full" charge profile which includes the effects of both disorder ($\epsilon$) and inter-electron Coulomb repulsion ($\phi$). However, to develop things further, we must separate these perturbative contributions out from those which remain in their absence. We therefore Taylor expand the above in powers of total disorder ($\widetilde{\epsilon}$) 
\begin{align}
    n_i 
    &  =
    \underbrace{ n_i |_{\widetilde{\epsilon}_i \rightarrow 0}  }_{ {n_0}_i = \langle n \rangle }  
    + \underbrace{ \left. \frac{ \partial n_i }{ \partial \widetilde{\epsilon}_i } \right|_{\widetilde{\epsilon}_i \rightarrow 0} \widetilde{\epsilon}_i }_{\delta n_i}
    + \mathcal{O}[\widetilde{\epsilon}^2].
    \label{eq:niClassicalTaylor}
\end{align}
Thus we find that, to linear-order in $\widetilde{\epsilon}$s, the local charge correction is given by 
\begin{align}
     \delta n_i  & = {n}_{i} - \langle n \rangle = A \widetilde{\epsilon}_i,
\end{align}
where $n_i$ was provided in (\ref{eq:niFermi}), while
\begin{align}
    \langle n \rangle 
    = {n_0}_i 
    = n_i |_{\widetilde{\epsilon}_i \rightarrow 0} 
    \overset{ (\textrm{ \ref{eq:niFermi}}) }{=}
    n_F(-\mu)
    = \frac{1}{e^{- \beta  \mu  } + 1}.
\end{align}
is now our classical version of the bare charge profile, and
\begin{align}
    A 
    & =   
    \left. \frac{ \partial n_i }{ \partial \widetilde{\epsilon}_i } \right|_{\widetilde{\epsilon}_i \rightarrow 0} 
    =  \left[ \frac{ \partial}{ \partial \widetilde{\epsilon}_i } \frac{1}{e^{\beta h_i} + 1} \right]_{\widetilde{\epsilon}_i \rightarrow 0}
    \beta \, \langle n \rangle \, ( \langle n \rangle - 1 ).
    \label{eq:coefficientAclassical}
\end{align}
is the only perturbative expansion coefficient we have left to consider in the classical limit.

\section{INCLUSION OF EXCHANGE-CORRELATION EFFECTS}

We take  $\hat{H} \rightarrow \hat{H} + \hat{V}^{\textrm{XC}} $, where
\begin{equation}
    \hat{V}^{\textrm{XC}} = \sum_i  \epsilon^{\textrm{XC}}_i \, \hat{c}^{\dagger}_i \hat{c}_i
    \qquad ; \qquad 
    \epsilon_i^{\textrm{XC}} = \epsilon_i^{\textrm{XC}}[n_i]
\end{equation}
within the local-density approximation made popular by standard density-functional methods. Noting further that,  with $\delta n_i \ll $ for all $i \in N$ as assumed in our perturbative scheme,
\begin{align}
    \epsilon_i^{\textrm{XC}}[n_i] 
    & = 
    \epsilon_i^{\textrm{XC}}[\langle n \rangle] 
    + 
    \underbrace{ 
    \left. \frac{ \partial \epsilon_i^{\textrm{XC}}[n_i] }{\partial n_i}\right|_{n_i = \langle n \rangle } }_{v^{\textrm{XC}}[\langle n \rangle]} 
    \delta n_i
    + \mathcal{O}[\delta n_i^2]
    \\
    & = \epsilon^{\textrm{XC}} + v^{\textrm{XC}} \delta n_i  + \mathcal{O}[\delta n_i^2]
\end{align}
can be truncated to first-order in $\delta n_i$. The zeroth-order term is evaluated with respect to the uniform solution (hence the absence of $\epsilon^{\textrm{XC}}$'s subscript in the second line, and similarly for $v^{\textrm{XC}}$) which provide a uniform shift to every site. It may therefore be absorbed into our chemical potential $\mu$. The first-order term, on the other hand, captures the exchange-correlation effects due to charge fluctuations $\delta n_i$ about the uniform result $\langle n \rangle$, providing our model with an additional source of perturbations. We thus revise our Hamiltonian to include these exchange-correlation effects  (compare with (\ref{eq:H}))
\begin{align}
    \hat{H} = \hat{H}_0 + \hat{V}
    \qquad ; \qquad 
    \left\{
    \begin{aligned}
        \\[-0.35cm]
        \hat{H}_0 
        & = \sum_{\vb{k} } \xi_{\vb{k}} \,  \hat{c}^{\dagger}_{\vb{k}}\hat{c}_{\vb{k}}
        \\
        \hat{V} 
        & = \sum_i (\epsilon_i + \phi_i + v^{\textrm{XC}} \delta n_i ) \, \hat{c}^{\dagger}_i \hat{c}_i 
    \end{aligned}
    \right.
    \label{eq:bareH}
\end{align}
Following the procedure outlined in \cref{subsec:PTgen} of the main text, we then obtain a linearized system of self-consistent equations which are Fourier transformed as follows (compare with (\ref{eq:FT_sysEqns}))
\begin{align}
    \left.
    \begin{aligned}
        {\delta n}_{i}   & = A \widetilde{\epsilon}_i + \sum_{j\neq i} B_{ij} \widetilde{\epsilon}_j 
        \\[0.1cm]
        \phi_i  & = \sum_{j \neq i} V^C_{ij} \delta n_j  
        \\[0.1cm]
        \widetilde{\epsilon}_i & = \epsilon_i + \phi_i + v^{\textrm{XC}} \delta n_i 
    \end{aligned}
    \right\}
    \quad
    \longrightarrow 
    \quad
    \left\{
    \begin{aligned}
        \delta n_{\vb{k}} & =  A \widetilde{\epsilon}_{\vb{k}} + B_{\vb{k}} \widetilde{\epsilon}_{\vb{k}}   \vphantom{ \sum_{j\neq i}  }
        \\[0.1cm]
        \phi_{\vb{k}} & = V^C_{\vb{k}} \delta n_{\vb{k}}
        \vphantom{ \sum_{j \neq i} V^C_{ij} \delta n_j  }
        \\[0.1cm]
        \widetilde{\epsilon}_{\vb{k}} & = \epsilon_{\vb{k}} + \phi_{\vb{k}} + 
        v^{\textrm{XC}} \delta n_{\vb{k}}     
    \end{aligned}
    \right.
    .
    \hspace{-0.1cm}
\end{align}
The $\vb{k}$-space solution is then given by (compare with (\ref{eq:δnk})) and (\ref{eq:φk}))
\begin{align}
    \delta n_{\vb{k}}
     & = M_{\vb{k}} \epsilon_{\vb{k}}
     \qquad ; \qquad 
      M_{\vb{k}} = \frac{A + B_{\vb{k}}}
      {1 - (A + B_{\vb{k}}) \left(V^C_{\vb{k}} + v^{\textrm{XC} } \right)}
    \\
    \phi_{\vb{k}}
    & = \widetilde{M}_{\vb{k}} \epsilon_{\vb{k}} 
     \qquad ; \qquad 
        \widetilde{M}_{\vb{k}} = V^C_{\vb{k}} M_{\vb{k}} 
\end{align}


\bibliographystyle{apsrev}
\bibliography{references}

\end{document}